# Ergodicity in Natural Earthquake Fault Networks

K. F. Tiampo[1], J. B. Rundle[2], W. Klein[3], J. Holliday[2], J.S. Sá Martins[4], & C. D. Ferguson[5]

[1] Department of Earth Sciences, University of Western Ontario, London, Ontario, Canada

[2] Center for Computational Science and Engineering, University of California, Davis, CA 95616, USA

[3] Dept. of Physics and Center for Computational Science, Boston University, Boston, MA 02215, USA.

[4] Instituto de Fisica, Universidade Federal Fluminense, Av. Litoranea s/n, Boa Viagem, Niteroi 24210-340, RJ, Brazil

[5] Council on Foreign Relations, 1779 Massachusetts Avenue, Washington, DC 20036, USA

Numerical simulations have shown that certain driven nonlinear systems can be characterized by mean-field statistical properties often associated with ergodic dynamics [C.D. Ferguson, W. Klein, and J.B. Rundle, Phys. Rev. E 60, 1359 (1999); D. Egolf, Science 287, 101 (2000)]. These driven mean-field threshold systems feature long-range interactions and can be treated as equilibrium-like systems with dynamics that are statistically stationary over long time intervals. Recently the equilibrium property of ergodicity was identified in an earthquake fault system, a natural driven threshold system, by means of the Thirumalai-Mountain (TM) fluctuation metric developed in the study of diffusive systems [K.F. Tiampo, J.B. Rundle, W. Klein, J.S. Sá Martins, and C. D. Ferguson, Phys. Rev. Lett. 91, 238501 (2003)]. In this work we analyze the seismicity of three naturally-occurring earthquake faults networks from a variety of tectonic settings in an attempt to investigate the range of applicability of effective ergodicity, using the TM metric and other, related statistics. Results suggest that, once variations in the catalog data resulting from technical and network issues are accounted for, all of these natural earthquake systems display stationary periods of metastable equilibrium and effective ergodicity that are disrupted by large events. We conclude that a constant rate of events is an important prerequisite for these periods of punctuated ergodicity, and that while the level of temporal variability in the spatial statistics is the controlling factor in the ergodic behavior of seismic networks, no single statistic is sufficient to ensure quantification of ergodicity. Specifically, we demonstrate that stationarity, while a necessary condition, is not sufficient to ensure ergodicity in fault systems.

## I. Introduction

Understanding the physics of earthquake faults and earthquake fault systems is essential for the development of possible forecasting methods for earthquakes or for showing that such methods are not possible. A class of forecasting techniques, which we will refer to here as statistical methods, have shown some promise for assigning reliable probabilities to large events (M > 5) over a decadal time period [4-6]. Although these techniques are promising, there are a number of unanswered questions associated with these methods. For example, they rely on linear decomposition techniques to quantify the temporal behaviour of a nonlinear, if slow-moving, system. In addition to this apparent dynamical contradiction, it also implies that the limitations on the forecast time period are poorly understood. A better understanding of the physics of earthquake fault systems is necessary in order to improve our ability to quantify its spatiotemporal dynamics.

Earthquake fault networks are a subset of a class of natural nonlinear systems, driven threshold systems [7-15]. In extensive simulations of fault networks formulated in such a way as to incorporate this threshold nature, the following picture is emerging: When the models have realistic long-range stress transfer [16, 17] they are in a state which we call punctuated or intermittent equilibrium near a critical point or spinodal [1, 3, 5-8, 18-22]. In this state the system appears to be in equilibrium for long periods of time. During these periods the statistical distribution of earthquakes is described by a Boltzmann factor where the temperature is fixed by an externally applied noise. Very large events in the models do not exhibit scaling. The scaling of events, referred to as Gutenburg-Richter scaling in real fault systems, is due, in the models, to the proximity to a spinodal.critical point [18, 19]. Moreover, these large events drive the system out of equilibrium for some period of time, after which the (quasi) equilibrium state is re-established. We refer to this as punctuated or intermittent equilibrium [18, 19].

A number of investigators have shown that these models of statistically stationary, driven dissipative mean-field systems appear to demonstrate effectively ergodic dynamics and these model systems reside in a sequence of physical states that are similar to equilibrium, or metastable equilibrium, states, punctuated by periods of instability, where equilibrium properties such as ergodicity do not exist [1-3, 8]. Recently, we showed that one particular naturally occurring system, an earthquake fault network, demonstrates the same punctuated equilibrium, as defined by effective ergodicity and measured using a fluctuation metric developed by Thirumalai and Mountain [3, 23, 24]. The purpose of this work is to investigate whether the punctuated equilibrium state also exists in other earthquake fault systems, and whether this punctuated equilibrium is a property of the earthquake process, or a function of the limits and accuracy of the seismic data and catalogs. To test this hypothesis we again employ this method, known as the Thirumalai-Mountain(TM) metric.

Although ergodicity is not sufficient to guarantee equilibrium, or the applicability of Boltzmann statistics, it is a necessary condition [1, 2, 25]. The establishment of this link between mean-field computational models of driven threshold systems and a natural earthquake fault system through the ergodicity property is important because it suggests that the long-term stationarity that is seen in these models also exists in the natural system. However, a number of questions remain. For example, the models display varying ranges over which ergodicity holds, apparently linked to the driving forces the system is subjected to in the simulation [2]. Here we investigate the variation in ergodicity for several different tectonic regimes, in an attempt to investigate whether the mechanics of the underlying fault structure affect the nature of punctuated equilibrium. We also provide a more detailed analysis of the effect of earthquake magnitude and spatial scale on the ergodicity property, in order to gain insight into their interrelationship.

The remainder of this paper is structured as follows: In section II we present a discussion of earthquake dynamics. In section III we introduce the TM metric and in section IV we discuss its application to fault systems. In section V we describe the fault systems we study and in section VI we present our results. Finally, in section VII we present our conclusions and discuss future work.

## II. Earthquake Dynamics

A key element of driven threshold systems is that their dynamics are strongly correlated in space and time over a multiplicity of spatial and temporal scales. In particular, if the range of interactions between elements of the system is long and the coupling weak, so that the dynamics can be understood as mean-field, fluctuations tend to be suppressed, but not eliminated, and the system may approach an equilibrium state (i.e., described by a Boltzmann factor) [16, 17]. In addition, Boltzmann fluctuations, which are an important property of equilibrium systems, have been directly observed in driven mean-field slider block simulations [7, 18-22]. Ferguson et al. showed, using a fluctuation metric, the Thirumalai-Mountain (TM) metric, originally developed to test for the presence of ergodic behavior in thermal systems [23, 24], that driven mean-field cellular automata versions of slider block models could also be considered to demonstrate ergodic behavior over finite intervals of time [1]. The same result was obtained for the Burridge-Knopoff model of earthquake systems [26]. Finally, direct observations of Gaussian fluctuations and detailed balance in transition probabilities provide evidence of ergodic behavior in a driven system of mean-field coupled map lattices [2].

One critical question is whether these conclusions from models of driven, dissipative systems can be extended to natural driven systems. We previously showed that the California earthquake fault system displayed ergodic behavior over a large enough spatial and temporal region, and that these periods are broken by large events that perturb the system away from equilibrium [3].

Driven nonlinear threshold systems are composed of interacting spatial networks of cells, each with one or more inputs, an internal state variable $\sigma(t)$ that evolves in time in response to inputs, and one or more outputs. Each cell is connected to an external driving source, and to the other cells by means of an interaction network. Threshold dynamics arises when a cell is subjected to this persistent external forcing that increases the value of $\sigma(t)$ through time until a predefined failure threshold $\sigma^F$ is reached, and the cell fails, reducing $\sigma(t)$ to a residual value $\sigma^R$. Thresholds, residual values, internal states, and the resulting dynamics are all modified by the presence of noise and disorder. Interactions between cells leads to dynamical self-organization in these systems, and may be excitatory (positive) in the sense that failure of connected neighbors brings a cell closer to firing, or inhibiting (negative) in the opposite case [6]. Mean-field threshold systems arise when the coupling, or interaction, between cells is long range but weak, leading to suppression of all but the longest wavelength fluctuations. These dynamics often result in strong space-time correlations in oscillator firings, along with the appearance of multiple scales in space and time.

In the mean-field regime, as the interaction length becomes large, resulting in a damping of the fluctuations, a mean-field spinodal appears that is the classical limit of stability of a spatially extended system. Examined in this limit, driven threshold systems display equilibrium-like behavior, including locally ergodic properties [18, 19, 27, 28]. Following the initial discovery that driven mean-field slider block systems with microscopic noise display equilibrium properties [1, 7], other studies have confirmed local ergodicity, the existence of Boltzmann fluctuations in both these and other mean- or near mean-field systems, and the appearance of an energy landscape as in other equilibrium systems [1, 3-8, 18-22].

In the 1990s, two major developments in seismic research greatly added to the understanding of earthquake fault network as a complex dynamical system. Stress transfer interaction studies demonstrated that fault interactions are in large part controlled by the stress state of the underlying geophysical medium [29-33]. On the other hand, a combination of theoretical analysis and numerical simulations established the link between earthquake fault networks and the physics of critical point systems, characterized by nonlinear dynamics, nonclassical nucleation, large correlation lengths, and the Gutenberg-Richter scaling law [1, 3, 8, 18, 19, 34-50].

As in other threshold systems, interactions in the natural earthquake system occur along a spatial network of cells, or fault segments, and are mediated by means of a potential that, in this case, is a function of the local and regional stress regime. The stresses are redistributed to other segments following slip on any particular segment, resulting in the diffusive nature of the stress field. For faults embedded in a linear elastic host, this potential is a stress Green's function whose exact form is calculated from the equations of linear elasticity. A persistent driving force, arising from plate tectonic motions, increases stress on the fault segments [1, 8, 18].

Once the stresses reach a threshold characterizing the limit of stability of the fault, a sudden slip event occurs. The slipping segment then may trigger slip at other locations on the fault surface whose stress levels are near the failure threshold at that time. In this manner, larger earthquakes result from the stress interactions as well as the persistent plate tectonics forces [1, 8, 47, 49]. The system dynamics are controlled, therefore, by this driven threshold mechanism overlain by the internal diffusion of stress on shorter spatial and temporal scales.

Because the external fault medium is elastic, the interactions between different patches and different faults are also elastic over short time scales. This fundamental elastic interaction results in the formulation of a mean-field regime for the earthquake system [1, 8, 47, 49]. This mean-field behavior occurs in elastic systems due to the inverse-cube nature of the stress Green's function (proportional to $1/|x-x'|^3$) that controls the elastic stress interactions between the fault patches in the medium [1, 8]. These long-range interactions lead to an averaging of stress over the system, damping the effects of short wavelength details [50]. Longer spatial and temporal wavelength effects become increasingly important, and the correlation lengths become increasingly larger as they approach a spinodal critical point, in association with power law scaling similar to the Gutenberg-Richter relation [1, 43, 50, 51].

The specific example used to represent these driven mean-field threshold systems is the slider block model for earthquake faults. Consider a two-dimensional (d=2) network of blocks sliding on a frictional surface, with the blocks arranged in a regular lattice pattern. Each block is connected to q other blocks by means of linear coupling springs, each having a spring constant $K_C$, and to a loader plate by means of a loader spring, again linear, having constant $K_L$. The loader plate translates at a fixed velocity V. In the simplest case of nearest-neighbor interactions, q = 2d, but when q >> 2d, and $qK_C \rightarrow$ constant, mean-field systems result [1, 8, 18]. For slider block models, the state variable $\sigma_i(t)$ is the force or stress on the $i^{th}$ block. The persistent motion of the loader plate raises the level of stress on all blocks over time. When $\sigma_i(t) = \sigma^F$, the block begins to slide, and comes to rest when $\sigma_i = \sigma^R$. The original Burridge-Knopoff slider block model, which had massive blocks, has recently been shown to display the ergodicity property when the springs between the blocks are long-range [26]. Other investigations emphasized the use of Stochastic Continuous Cellular Automaton (SCCA) models, in which the blocks are massless, the motion is over-damped, and in which the sliding block is subjected to an additional small-amplitude random force that plays the role of a thermalizing noise [1, 8, 18]. In these SCCA models, frequently used to model earthquake fault systems, ergodic properties are observed as well [1, 3, 8].

In these numerical simulations, mean-field earthquake networks are nonequilibrium systems that can exhibit properties of equilibrium systems as they settle into a metastable equilibrium state. The time averaged elastic energy of the system fluctuates around a constant value for some long period of time. These periods are punctuated by major events which reorder the system before it settles into another metastable well around a new mean energy state [1, 3, 8, 50]. During these reorderings the system is not ergodic; hence, we refer to these systems as punctuated ergodic.

Note that the spatial and temporal firing patterns of such driven threshold systems develop from the obscure underlying structures, parameters, and dynamics of the multidimensional nonlinear system [52]. As a result, these patterns are complex and often difficult to understand and interpret from a deterministic perspective. For example, while it is not possible to measure all the cellular potentials of the neurons of the human brain and the physical and chemical parameters which control the temporal evolution of its potentials and currents, it is possible to observe the complex firing patterns of neural cells [53, 54]. Similarly, there is no means at present to measure the stress and strain at every point in an earthquake fault system, or the constitutive parameters that characterize this heterogeneous medium and its dynamics, but those patterns that express themselves in the resulting seismicity can be observed from the surface of the earth [55-60].

This seismicity, the firing patterns that are the surface expression or proxy for the dynamical state of the underlying fault system, can be located in both space and time with considerable accuracy [61-63]. If this natural system is, as simulations suggest, a mean-field threshold system in punctuated metastable equilibrium [1, 8, 50], then the time averaged elastic energy of the system fluctuates around a constant value for some period of time. These periods are punctuated by major events that reorder the system before it settles into another metastable energy well. In addition, if the system behaves ergodically for significant periods of time, then it is stationary for those same periods. Recent studies [5, 6] suggest that linear operators can effectively forecast the dynamical behavior of these systems with certain limitations. It is also known that the ergodicity is a necessary condition for the effective use of these linear, Karhunen-Loeve operators, in order to ensure that it has defined a complete, orthonormal set of basis vectors [25, 64]. As such, it is important to quantify the limits and conditions under which ergodicity exists in these fault systems in order to better understand their dynamics and applicability. In addition, while the system may be ergodic for long periods of time, if large events can be directly identified that drive the system away from ergodicity, this supports the work of others suggesting that large earthquakes are equilibrium-breaking events. Here we will employ a quantity called the Thirumalai-Mountain (TM) fluctuation metric in order to study the possible existence of punctuated ergodic properties of several natural earthquake fault systems [1, 3, 23, 24].

## III. The Thirumalai-Mountain Metric

The TM metric measures effective ergodicity, or the difference between the time average of a quantity, generally related to the energy, $E_j$, at each site, and its ensemble average over the entire system. The fundamental idea is that of statistical symmetry, in which the N oscillators, particles, cells, or spins in the system are statistically identical, in terms of their averaged properties – the statistics of one particle look the same as the statistics for the entire system [23, 24, 65]. While most systems are ergodic for infinite averaging times, if the actual measurement time scales are finite, but long, a large sub-class will have all regions of phase space sampled with equal likelihood and the system is effectively ergodic [24]. Exceptions to this are systems such as glasses, which are ergodic for infinite averaging times but for finite averaging times are not. Practically, for *effectively* ergodic systems, the spatial and temporal averages are constant over a large enough representative sample in time and space. Ergodicity is a behavior that is generally limited to equilibrium states, in which transition probabilities are univarying or follow a definite cycle, and implies stationarity as well. Therefore, if such a system is ergodic, it is also in some form of stable or metastable equilibrium and can be analyzed as such.

The fluctuation metric $\Omega_e(t)$, proposed by Thirumalai and Mountain is

$$\Omega_e(t) = \frac{1}{N}\sum_{i=1}^{N}[\varepsilon_i(t) - \bar{\varepsilon}(t)]^2, \quad (1)$$

where

$$\varepsilon_i(t) = \frac{1}{t}\int_0^t E_i(t')dt' \quad (2)$$

is the time average of a particular individual property, $E_i(t)$, and

$$\bar{\varepsilon}(t) = \frac{1}{N}\sum_{i=1}^{N}\varepsilon_i(t) \quad (3)$$

is the ensemble average of that temporal average over the entire system. Here $\Omega_e(t)$ is the value of the TM metric computed at increasing time *t*, and the start of the catalog corresponds to the initial time, *t* = 0. $E_i(t)$ is a physical measure that represents the value of the energy of the $i^{th}$ particle at time *t*. If the system is effectively ergodic at long times, $\Omega_e(t) = \frac{D_e}{t}$, where $D_e$, the ergodicity diffusion parameter, is related to the rate at which the phase space is explored, and proportional to the inverse of the time scale needed to reach ergodic [23, 24]. Physically, the deviation of the time-averaged quantity from its ensemble average in equation (1) is decreasing as a function of time, and all particles in the system are statistically similar.

Note, also, from above, that the TM metric is actually a measure of the spatial variance of the temporal mean over a time interval *t* at each location, calculated at each successive time step. As a result, if the dynamics are sampling all of the phase space equally, and there is a general equivalence of the temporal and ensemble averages, then the central limit theorem holds such that the variance, which becomes a constant controlled by the large sample size *N*, is divided by the increasing time, *t* [23].

In slider block models used to replicate the behavior of earthquake fault networks, as the interaction range increases, the system approaches mean-field limit behavior. If, as a result, these slider block models are in metastable equilibrium, they can be analyzed using the methods and principles of equilibrium statistical mechanics. Ferguson et al. [1] applied the TM metric to the energy of each block in slider block numerical simulations to show that the system was ergodic at external velocities, V, that approach V = 0. The data showed the expected increasing linear relationship between the inverse TM metric and time such that

$$\frac{1}{\Omega_e(t)} = \frac{t}{D_e}, \qquad (4)$$

denoting effective ergodicity as defined above and as it is typically represented [1, 23].

For a similar slider block model [4], the TM metric was calculated not for the energy but for numbers of events. In this calculation of the TM metric, $E_i(t) \equiv R_i(t)$, the number of events greater than a certain magnitude. Note that number of events is a proxy for energy release, as detailed below. In this case, there was an initial transient phase, in addition to the linear sections expected from equation 4, where the system exhibits ergodic behavior punctuated by the occurrence of larger events [3].

Not only have slider block models been shown to be in metastable equilibrium, and can be analyzed as such [1, 8], we demonstrated previously that the same applies to at least one natural earthquake fault network [3] and that the interactions in a natural driven system are also mean-field in the ergodic case. Here we proceed to study different types of regional earthquake networks by applying the TM fluctuation metric to the associated fault system seismicity.

## IV. Application

We apply the TM metric to the surface expression of the energy release in a regional fault system. For application to the earthquake fault system, the number of earthquakes of a particular magnitude or greater can be expressed as a function of the seismic energy release. If *N(t)* is the seismicity rate, or number of events for a given time period, for earthquakes of magnitude greater than *m*, and *a* is the rate for all events over a given region, then the Gutenberg-Richter law can be expressed as

$$N(t) = a10^{-bm}, \qquad (5)$$

or $$\log N(t) = a - bm, \qquad (6)$$

where $b \approx 1.0$, and $a$ is a constant over the region of interest [55, 66, 67]. Also, if $E(t)$ is the energy for a given event magnitude $m$, in joules, $c \approx 1.44$ and $d \approx 5.24$ [66, 67], and

$$E(t) = 10^{cm}10^{d}, \tag{7}$$

such that

$$m = \frac{\log E(t) - d}{c}. \tag{8}$$

Substituting,

$$\log N(t) = a - \frac{b(\log E(t) - d)}{c} = a' - b' \log E(t), \tag{9}$$

and

$$N(t) = 10^{a'} E(t)^{-b'}, \tag{10}$$

where $a'$ and $b'$ are again constants that depend on the region and time period of interest. Typically, $b$ is approximately 1.0, and $a$ represents the background rate of seismicity for the particular region and a given minimum magnitude [55, 66, 67]. The number of events in a given time period, $N(t)$, greater than a given magnitude, $m$, is therefore a function of the seismic energy release.

For this particular application of the TM metric, we used $E_i(t) \equiv N_i(t)$, the number of events greater than some minimum magnitude $m$, calculated for each year. Specifically, we specify for each region or subregion under analysis in each tectonic zone (California, Spain, or Canada) a grid, or set of boxes, that we will use to separate our seismic events by location, designated by $i$. The boxsize is varied as we attempt to investigate its effect on the resulting ergodic properties of the region (see below), but is most often set to a default value of 0.1° in the latitude and longitude directions.

For each box we count the number of seismic events to occur during a specific time period which is, for all subsequent analyses discussed here, at one year. This is our seismicity rate, $N_i(t)$, or number of events for each year at each location $i$. Again, for our purposes in calculating the TM metric and its inverse, $N_i(t)$ is a proxy for energy, and substituted for $E_i(t)$ in equation (2), above.

## V. Seismicity Data

For this analysis, we employed catalogs from three different regions of the world – California, Spain, and eastern Canada. These three particular catalogs were chosen because they represent very different tectonic regions, but the seismicity in each of these regions is shallow, limiting the need to account for a depth effect, and they are each well characterized. Here we include additional TM analyses of the seismicity catalogs for the Iberian peninsula and eastern Canada as well as a more comprehensive TM analysis of the California data than that included in our initial work [3], for a variety of magnitude cutoffs, time periods and spatial scales, and an investigation into the constitutive features that underlie these periods of effective ergodicity.

### A. California

The California fault system consists of a large transform boundary between the Pacific and North American plates that spans almost 1100 kms. The San Andreas, a right-lateral strike-slip fault capable of producing events as large as M~8, is the dominant structure; however, the 15-to-20 km deep elastic region is broken by a variety of faults of different sizes and mechanisms, producing a variety of patterns and behaviors throughout the region [62]. Figure 1 shows a schematic of the California fault network, superimposed upon which are the largest events to have occurred in the region since 1932.

The seismicity data employed in our analysis is taken from existing observations in California between the years 1932 and 2004, and is an amalgamation of data from the Southern California Earthquake Center (SCEC) (www.scec.org) and the Northern California Seismic Network (NCSN) (quake.geo.berkeley.edu). Note that, while the network coverage has changed significantly since its inception in the early 1930s, the catalog is not declustered in any way. Using various subsets of this data covering the period from January 1, 1932 through December 31, 2004, we compute the TM metric for California seismicity, over the region 32º to 40º latitude, -115º to –125º longitude.

### B. Eastern Canada

Eastern Canada is an intraplate, shield region in which most of the earthquake activity is associated with large lithospheric-scale tectonic and geological structures arising from past orogenic and rifting episodes, and that appear to control the spatial distribution of seismicity [68]. Figure 2 shows the largest events to occur in the region since 1900. Activity varies from low background seismicity to medium (magnitude ranging from 4 to 6) and large (magnitude M~7) earthquakes, and the bulk of the seismicity occurs within the top 25 km of the crust. Here we compute the TM metric for the region 42º to 52º latitude, -60º to -85º longitude.

C. Spain

The Iberian Peninsula is complicated and diffuse, controlled to the south by the convergence between the Eurasian and African plates, but it is also affected by a general uplift causing radial extension [69]. As a result, the regional seismicity of the Iberian Peninsula is complicated, characterized by a diffuse geographical distribution that ranges from low to moderate magnitudes, with a maximum depth of 146 km, although the bulk of the seismicity is shallower. Figure 3 shows the largest events to occur in the region since 1970. Earthquakes rarely exceed magnitude 5.0 either on the Iberian Peninsula itself or in northern Morocco, the westernmost Mediterranean Sea and the Atlantic Ocean south of Portugal [69].

The data used has been recorded by the Geographic National Institute, which runs the National Seismic Network with 42 stations, 35 of them of short period, connected in real time with the Reception Center of Seismic Data in Madrid. The catalog, with more than 10000 earthquakes in the region between 35º north and 45º latitude and between -5º to 15º longitude, contains all the seismic data in the Iberian Peninsula and northwestern Africa collected in the period 1970 - 2001.

## VI. Results

A. California

In Figure 4 we plot the inverse TM metric for the number of events in southern California, from 1932 through 2001. The linear relationship between the inverse TM metric and time can be observed again, as seen in the numerical simulations and in our previous work with California seismicity [1, 3, 8]. Note that Figure 4 is a correction to Figure 3 of [3], where the boxsize was incorrect for events of magnitude three. Here, the system is determined to be ergodic when the inverse TM metric is linear, with a positive slope. Note that, for a boxsize of 0.1 and $M \geq 2$, the California earthquake fault system is ergodic for only a short period of time between approximately 1955 and 1968 (Figure 4a). It is ergodic for M ≥ 3 from 1932 to approximately 1970, but not for the subsequent time period (Figure 4b). The fault system is ergodic over long stretches for the time period shown in Figure 4c, M ≥ 4.

Several interesting conclusions can be drawn from Figure 4. While in our earlier work [3] we studied only events of M ≥ 3, here we show analyses for three different cutoff magnitudes. The results for M ≥ 2 are not intuitive (Figure 4a). Ergodicity implies a condition in which all regions of phase space are sampled with equal likelihood. While it might be hypothesized that by adding more small events the system would become more ergodic, sampling the phase space more evenly, this is not the case, confirming that ergodicity is not related directly to the number of events, or an outcome of random sampling of many events. Note that there are approximately six times as many events of M ≥ 2 as M ≥ 3. One potential explanation for this result is that, by adding many small events that occur over the entire network, the system is sampling one portion of the phase space, the background seismicity, more preferentially. Another factor may be that, as the catalog is not complete for M ≥ 2 over the entire time period, the increasing ability to sample more events over time with better instrumentation and coverage affects the phase space coverage. We will investigate these possibilities below.

Figure 4b shows the inverse TM metric for M ≥ 3. Here we see that the system is ergodic for long periods prior to the early 1970s, and is then ergodic for short periods throughout the 1970s, but is no longer ergodic after 1979. We hypothesize that the early ergodic behavior is related to the stability of the networks and catalog prior to 1970, when digitization of the seismic networks began and many more events were recorded in new spatial regions [70, 71].

For events of M ≥ 4, the California fault system is effectively ergodic for relatively long periods of time, on the order of decades, from 1932 through 2004. These equilibrium periods are punctuated by the occurrence of large earthquakes, shown in Table 1 and highlighted with arrows on Figure 4c. These include the Kern County event of 1952, the Imperial Valley earthquake in 1979, the Landers sequence of 1992, and the Hector Mine earthquake in 1999. Between these events the fault system resides for long periods in metastable equilibrium, evidenced by this punctuated ergodicity. Eventually, the nonlinear dynamics lead to a large earthquake, and the system departs from its current local energy minimum, to migrate to a new local minimum, where it again resides in an effectively ergodic state.

In Figure 5 we plot the number of events per year for different magnitude cutoffs. Here it is useful to detail several known issues with the historic catalogs, and several additional constraints uncovered during our recent investigations. First, spatial resolution varies with time. Before 1963, locations in the ANSS catalog are binned to the arc-minute (or 1.83 km), while after 1965 they are binned to the decimal degree. The SCEDC catalog has a nearly uniform spatial distribution throughout the entire time period, although there is a slight preference for events to be recorded on the whole degree (rather than some decimal degree) before 1965.

The catalog completeness is a related issue. Until the 1950s, the Gutenberg-Richter distribution curves shows roll-off at around M ~ 3.5. The 1960s are complete for M ≥ 3.0 (roll-off starting at M~3), while the 1970s until the present appears complete down to M ~ 2.

For M ≥ 2 (Figure 5a), the seismicity rate remains relatively constant through the late 1960s, and increases sharply at the same time that the system becomes non-ergodic, as shown in Figure 4. Note that this coincides with digitization of the seismic network in the early 1970s. This increased the ability to detect smaller events, including the spatially and temporally clustered aftershocks of larger events. Note that it also coincides with the time period in which the catalog became complete for those magnitudes. It suggests that completeness is not a requirement for ergodicity, but that catalogs which are not complete can be stationary. It is also clear from Figure 5a that the number of events per year never approaches a constant mean or variance, accounting for the decrease in the inverse metric after 1968. It should be noted that this result is consistent with data supplied by ANSS [71], where the number of events after 1970 shows a constant rate of increase through the 1990s. From Figure 5, it is apparent that this steady increase in numbers of recorded events is primarily due to the increased detectability of M ≥ 2. However, this does not account for the lack of ergodicity prior to the early 1970s, implying that, while a requirement, a constant rate of events is not a sufficient condition for ergodicity, and supports our hypothesis that the record of events for M ≥ 2 is not sampling the phase space equally.

Figure 5b shows the number of events for M ≥ 3. Here the rate of events is approximately constant prior to the mid-1970s, and the catalog is effectively complete after 1960. This corresponds to a period of ergodic behavior, prior to an increase in both the yearly rate at the same time as the onset of the decrease in the inverse metric (Figure 4b) in the late 1970s.

For M ≥ 4, Figure 5c reflects a relatively constant mean value of the seismicity rate over the entire time period. Here the catalog is complete throughout the entire period for M ≥ 4, while the yearly rate remains constant. Increases are seen for short periods during and after the 1952 Kern County and 1992 Landers earthquakes, which correspond to deviations from ergodicity in Figure 4c. This again leads to the conclusion that while a constant mean rate of seismicity is important for the ergodic condition, spatial and temporal changes in the variance that result in non-ergodic behavior are important as well.

In the following figures, we investigate the behavior of those quantities that are used to calculate the TM metric as originally described, where every box is included in the calculation, even if it contains no events during the time period of interest, to match the conditions in our earlier figures. Figure 6 details the mean and variance in number of earthquakes, at a boxsize of 0.1 degrees, for cumulative magnitude, $M \leq m$. These values approach a constant over both time and space for cumulative magnitudes greater than 3.0. This reinforces the conclusion that the system is stationary over large enough spatial areas and time periods, as is implied by the ergodicity constraint. However, it also illustrates that stationarity is not a sufficient condition for ergodicity, as for a number of these spatial and temporal regions, at particular magnitude cutoffs, the system does not display ergodicity for significant periods of time (Figure 4). It should also be noted that the deviations between magnitude three and four are much more significant in the temporal statistics.

In Figure 7 is shown the spatial variance for California seismicity over time, for events of magnitude greater than or equal to that shown for different magnitude cutoffs and boxsizes. Figure 7a illustrates that, for $M \geq 3.0$ and a boxsize of 0.1 degree, the slope of the spatial variance changes significantly in the late 1970s, with the digitization of seismic records, corresponding to the breakdown of ergodicity seen in Figure 4b. The rate of change of the variance is nearly 10 times higher in the last 15 years than in the first 40, but is nearly constant for the period after 1980.

In Figure 7b, for a boxsize of 0.02 degree (~ 2.2 km) and $M \geq 3.0$, the rate of change of the variance over time is significantly less than 1 and approximately equivalent in the early time period, prior to 1950, and after 1980, although the slope shows some variation after the 1952 Kern County earthquake. Note also that the slope in the spatial variance changes noticeably in the early 1960s, when the locations changed from being binned on the arc-minute, a value approximately that of the discretization boxsize at 0.02 degrees, to decimal degrees. Finally, for $M \geq 4.0$ and a boxsize of 0.1 degree (Figure 7c), the slope is effectively constant, with a change per year that is still significantly less than one. The short-term divergence from ergodicity seen after large events (Figure 4c) corresponds to sudden increases seismicity due to aftershocks and a resulting increase in the variance after the 1952 Kern County, the 1979 Imperial Valley, the 1992 Landers, and the 1999 Hector Mine earthquakes.

In Figure 8 we illustrate the spatial variation of California seismicity for the same three magnitude ranges over the period 1932 to 2004. Here circle size increases with increasing magnitude. The first feature that one notices is the similarity between the three images. While there are considerably more events for M ≥ 2 (Figure 8a), the bulk of the seismicity occurs in the same regions as for M ≥ 3 (Figure 8b), reinforcing the conclusion that an increase in numbers of events (from magnitude three to magnitude two) does not necessarily result in ergodic behavior, as demonstrated in Figure 4. Figure 8c details the spatial distribution of seismicity, M ≥ 4, for 1932 to 2004. While the clustering is more localized in this figure, the locations of maximum activity remain the same as those shown in Figures 8a and 8b, suggesting that temporal variations in the spatial distribution must play an important role in the ergodic behavior.

Figure 9 shows the seismicity distribution, M ≥ 3, for the time period 1970 to 2004. The spatial variation for this figure is more like that of Figure 8c than Figure 8b. The similarity to the spatial variation for the M ≥ 4, which demonstrates punctuated ergodicity (Figure 4c), prompted the analysis of the seismicity for M ≥ 3 since 1970.

Figure 10 shows the inverse TM metric for a boxsize of 0.1 degrees, 1970 to 2004, M ≥ 3. The seismicity for this time period and discretization is effectively ergodic for the time periods between large events. Note that, for this time period, the tendency to discretize the spatial locations on the arc-minute, seen prior to 1965, as well as the lack of completeness for events of M ≥ 3 are no longer a factor. In this case, after an initial transient period, the system becomes ergodic after the 1983 Coalinga earthquake, the 1992 Landers sequence, and the 1999 Hector Mine event. Again, the pattern of seismicity seen for M ≥ 3 during this time period is remarkably similar to that seen for M ≥ 4 over the entire history, yet the total number of events is much larger. Note that it is the same large events that perturb the system away from ergodic behavior for both analyses.

Figure 11 shows the results of two additional TM calculations for M ≥ 3. Figure 11a shows the same time period as Figure 4b, 1932 through 2004, but again a boxsize of 0.02 degrees (approximately 2.2 km square). Figure 11b is the inverse TM metric, again for the period 1932 to 2004, but now for a boxsize that approximates the rupture dimension of an earthquake of magnitude three, 0.0011 degrees or 0.12 km [72]. In addition, for Figure 11b, the TM calculation is performed such that those boxes which contain no events are not included in the calculation, where $N$ is replaced by the number of boxes that are not empty in equations 1 and 3, above.

For a boxsize of 0.02, M ≥ 3, the system is effectively ergodic for the entire time period, except for a nonlinear section between approximately 1955 and 1965 (Figure 11a), coincident with the occurrence of the 1952 Kern County event and the discretization of the catalog at the same spatial scales as the boxsize. Note that here those periods in which the system appears to behave ergodically, but with different slopes, correspond to the different time periods shown in Figure 7b. There is a jog in the slope in 1971, coincident with the occurrence of the San Fernando earthquake and the digitization and expansion of the networks. Finally, the shorter-term breaks in ergodicity after 1980 appear to correspond to those seen in Figure 10, but with a decrease in the overall slope. The difference between this result and Figure 4b supports the proposition that a relationship exists between the boxsize, or the sampling rate of the total number of events, and the ergodic behavior. One possible interpretation is that if the sampling rate inherent in the boxsize is too low for the large numbers of events in the catalog (29,307 events of $M \geq 3$) the central limit theorem (CLT) no longer holds, and the spatial variance of the temporal mean no longer goes as $1/t$, as required for linearity in the TM metric (Equation 2). As the number of boxes, $N$, increases with smaller boxsize, the law of large numbers reasserts itself. If the random sample comes from a large population with a common distribution, and the variance is finite, the sample variance converges to the variance of the distribution divided by the sample size. In this case, the sample size is the number of years, and the TM metric, as the temporal variance of the spatial mean, is divided by $t$, the sample size, and which increases each year, creating the linear relationship. Physically, the smaller size of the boxes allows the seismicity to sample a larger number of individual realizations of the phase space with equal likelihood, reducing clustering and generating a common distribution, producing effective ergodicity.

Figure 11b displays ergodic behavior for the time period after 1980, for events of $M \geq 3$, but not before. This result supports that seen in Figures 4 and 10, but illustrating that the defects in the catalog discussed earlier are the dominant effect prior to 1980, and that the linear, ergodic behavior subsequent to that is as shown in Figure 10. The inset shows the results for 1980 to 2004, highlighting the departures from ergodicity as a result of earthquakes such as the 1992 Landers sequence. This result is supported by the work of Xia and colleagues [personal communication, 73], who found that the inclusion of large numbers of locations with zero events in calculating the TM metric for large-scale numerical earthquake simulations distorted the results. They began with a simulation based upon the Olami-Feder-Christensen model [74], and fixed a certain number of boxes such that there were no events allowed in those locations. The resulting inverse metric calculation was not linear, implying that the system was not ergodic. Removing those boxes with no events from the calculations caused the inverse TM metric to become linear again, correctly validating its ergodic behavior. Again, as stated above, here the boxsize is equivalent to the approximate rupture dimension for an earthquake of magnitude three [72], the lower magnitude cutoff for this analysis, and no boxes that without events are included in the calculation.

B. Spain

The inverse TM metric for the Iberian Peninsula is plotted in Figure 12, for a boxsize of 0.1 degrees, $M \geq 3.0$ (Figure 12a) and $M \geq 4.0$ (Figure 12b), 1970 to 2001. Here we see, again, that while the system is not ergodic for $M \geq 3.0$ at this boxsize, it does achieve ergodicity for $M \geq 4.0$, even for this relatively short time period.

In Figure 13 we show the spatial distribution of events from the Iberian catalog, where increasing circle size again denotes increasing magnitude, from $M \geq 3.0$ (Figure 13a) to $M \geq 4.0$ (Figure 13b). As in the case of California, the clustering in the seismicity is similar for both magnitude cutoffs, suggesting the variability in the temporal statistics is the controlling factor in the ergodic behavior of seismic catalogs.

Figure 14 plots the total number of events per year for the Iberian catalog, for both $M \geq 3.0$ (Figure 14a) and $M \geq 4.0$ (Figure 14b). Again the number of events per year does not remain constant for $M \geq 3.0$, in the case for the entire time period, 1970 to 2001. On the other hand, the seismicity rate for $M \geq 4.0$ remains effectively constant for the entire period, except for a minor variation in the early 1980s, at the same time as the system diverges from ergodicity in Figure 12b.

C. Eastern Canada

In Figure 15 we plot the inverse TM metric for the eastern Canadian catalog. This catalog covers the longest time period of all three catalogs, 1900 to 2001. The inverse TM metric is shown for a magnitude cutoff of $M \geq 3.0$ (Figure 15a) and $M \geq 4.0$ (Figure 15b).

Figure 15a, $M \geq 3.0$, displays ergodic behavior for the first 80 years, after an initial transient period, followed by a deviation from ergodic behavior beginning around 1980. As in the case of the California example, increased data collection and network coverage in eastern Canada in recent years and the associated change in earthquake statistics is the most likely explanation for this deviation from ergodicity [75]. One other possible explanation for the flattening that occurs after 1980 is the inclusion of boxes with zero events in this calculation, at a boxsize significantly greater than the rupture dimension of the minimum fault size, as illustrated for California (Figures 4 and 9), although additional testing is necessary to verify this. Figure 15b, $M \geq 4.0$, again displays ergodic behavior for the entire time period, suggesting that the temporal variation in the phase space configuration for that magnitude range is stable.

Figure 16 details the spatial distribution of earthquakes in eastern Canada for the same time period and magnitude cutoffs, where increasing circle size again denotes increasing magnitude. The spatial clustering for M ≥ 3.0 (Figure 16a) and M ≥ 4.0 (Figure 16b) is very similar, as in the case of the Spanish and California catalog, despite the lower number of events.

Figure 17 is a plot of the number of events per year for eastern Canada, M ≥ 3.0 (Figure 17a) and M ≥ 4.0 (Figure 17b). While the number of events per year for M ≥ 3.0 increases over the catalog life, the period for 1920 through 1980 displays a constant mean, corresponding to the ergodic behavior in Figure 15a, while the rate increases continuously subsequently to 1980. Figure 17b shows that the seismicity rate for M ≥ 4.0 has a relatively constant statistics for the entire period. In addition, it suggests that ergodic behavior is not dependent on large numbers of events, as the rate of M ≥ 4.0 earthquakes per year over the entire regions is always less than eight.

## VII. Conclusions

In conclusion, we employ here the Thirumalai-Mountain fluctuation metric and data from existing seismic monitoring networks to identify the presence of punctuated ergodicity, an equilibrium property, in the dynamics of the natural earthquake fault system in three varied tectonic regions; California, Spain, and eastern Canada. While the results are clearly impacted by the quality of the catalogs, in particular their magnitude of completeness and spatial discretization, if these effects are removed, as shown in all regions for M ≥ 4, and for M ≥ 3 in California (Figures 10 and 11), natural earthquake fault systems are effectively ergodic and mean-field for significant periods of time, as in the numerical simulations that are used to study these systems. Recent work in the study of aftershock sequences also concludes that the lower magnitude cutoff and spatial discretization significantly affects the statistical behavior of seismicity [76]. It is equally important to note that, once the effects associated with the data collection itself are removed, the large earthquakes still continue to perturb the system away from ergodicity, sending it from one metastable well to another dynamical state. The fact that it is ergodic, and can be measured as such by a technique designed to measure ergodicity in diffusive systems, means that it is both stationary and in equilibrium, if only temporarily, for those time periods. Divergence from ergodicity occurs when it is no longer in equilibrium and new areas of phase space are being explored.

All three of these fault systems display punctuated ergodic behavior for some combination of magnitude and boxsize that is clearly related to the occurrence of large events, despite that fact that one is located in a region of high seismicity and strong directional tectonic forcing (California), the second is a region of moderate seismicity and varied stress directions (Spain), and the third lies in an area of low seismicity and low intraplate stresses (eastern Canada). This work supports the general conclusion that natural fault systems display at least some of the dynamics of driven mean-field systems, as seen in numerical simulations of interacting slider blocks and coupled map lattices. These systems reside in metastable wells for significant periods of time for particular magnitude regions and spatial discretization (boxsize), on the order of several decades for all three fault systems. As the dynamical systems evolve, they migrate to a new free energy minimum with the occurrence of a large earthquake.

In a detailed study of the magnitudes and boxsizes applicable to ergodic behavior in each of these systems, we have determined that a lower magnitude cutoff of 4.0 appears to display ergodicity for all spatial and temporal regions. In addition, lower cutoff ranges require smaller discretization in order to ensure central limit theorem behavior and a $1/t$ increase in the inverse TM metric (Figures 10 and 11). We have also shown, by removing those locations from the calculation that do not contain any events over the time period (Figure 11) and employing a boxsize that corresponds to the rupture dimension of the lower magnitude cutoff, that it is the fault systems themselves that have the ergodicity property, and not geographical regions. Finally, earlier and shorter temporal regions appear to be affected by the quality and completeness of the catalogs.

We have performed a variety of statistical tests on the catalogs in an attempt to better understand the underlying ergodic behavior and its genesis. While it appears that a relatively constant rate of events is an important prerequisite, and that the level of temporal variability in the spatial statistics is the controlling factor in the ergodic behavior of seismic catalogs, no one statistic is sufficient to ensure quantification of ergodicity. We also show that stationarity, while a necessary condition, is not sufficient to ensure ergodicity in seismic networks. The TM metric, a measure of the spatial variance of the temporal mean, alone provides the ability to quantify ergodic behavior.

As noted earlier, if the system behaves ergodically for significant periods of time, then it is stationary for those same periods. Recent studies [4, 5] suggest that linear operators can effectively forecast the dynamical behavior of these systems with certain limitations. Because ergodicity is a necessary condition for the effective use of these linear, Karhunen-Loeve operators, in order to ensure that it has defined a complete, orthonormal set of basis vectors [25, 64], it is critical to quantify the limits and conditions under which ergodicity exists in these fault systems in order to better understand their potential applicability. Here we have demonstrated that the TM fluctuation metric can be employed to identify the spatial and temporal parameters for which punctuated ergodicity holds in natural earthquake fault systems in order to appropriately apply these and other, related techniques.

This work supports the appropriate application of the principles of equilibrium statistical mechanics to the study of the physics of earthquake fault systems from different tectonic settings, and provides a means of identifying their range of applicability. Through the application of the TM metric to various tectonic regions, we have demonstrated both its widespread applicability and that punctuated ergodicity can be identified in many natural fault systems, despite spatial and temporal variability in their underlying dynamics. The success of the TM metric for this application, a measure initially based upon the model of a Brownian particle in a diffusive system, also supports the interpretation of the natural earthquake system in which stress diffusion plays an important role in the dynamics, despite the variety of driving forces related to differences in the plate tectonic settings. Work is currently underway to employ the metric in order to better quantify the diffusive nature of the fault system and its dynamics in the future. Finally, we believe that, by accounting for the contribution of various effects related to data collection that are not a function of the underlying physics, we have shown that while natural fault systems are ergodic for long periods of time, large events can be directly identified that drive the system away from ergodicity. This supports the work of others suggesting that large earthquakes are equilibrium-breaking events. Future work will investigate the implications of these results as well.

Acknowledgements: Research by KFT was funded by an NSERC Discovery Grant. This research also was supported by the Southern California Earthquake Center. SCEC is funded by NSF Cooperative Agreement EAR-8920136 and USGS Cooperative Agreements 14-08-0001-A0899 and 1434-HQ-97AG01718. The SCEC contribution number for this paper is 813. Research by JBR was funded by USDOE/OBES grant DE-FG03-95ER14499 (theory), and by NASA grant NAG5-5168 (simulations). Research by WK was supported by USDOE/OBES grant DE-FG02-95ER14498 and W-7405-ENG-6 at LANL. WK would also like to acknowledge the hospitality and support of CNLS at LANL. CDF completed his part of this research while a graduate student at the Dept. of Physics, Boston University and received funding through WK's grants. The manuscript was significantly improved by the comments of three anonymous reviewers.

Special thanks goes to Dr. John Adams of the GSC, Canada and Dr. Antonio Posadas and Dr. Abigail Jiménez of the Department of Applied Physics, University of Almería, Spain for supplying the eastern Canadian and Spanish catalogs, respectively.

Correspondence and requests for materials should be addressed to K.T. (e-mail: ktiampo@uwo.ca).

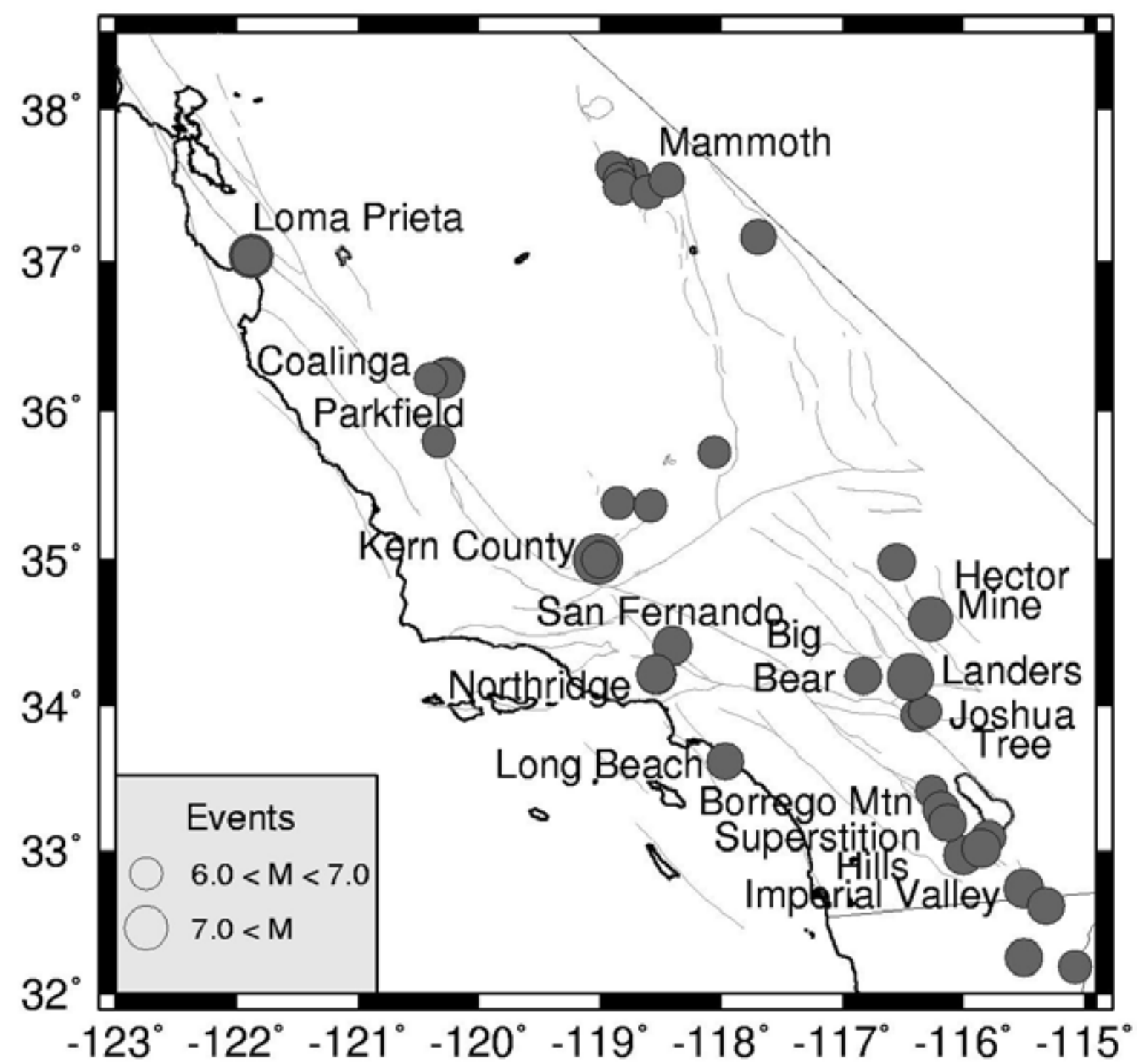

FIG. 1: Large events in southern California, 1932-2004.

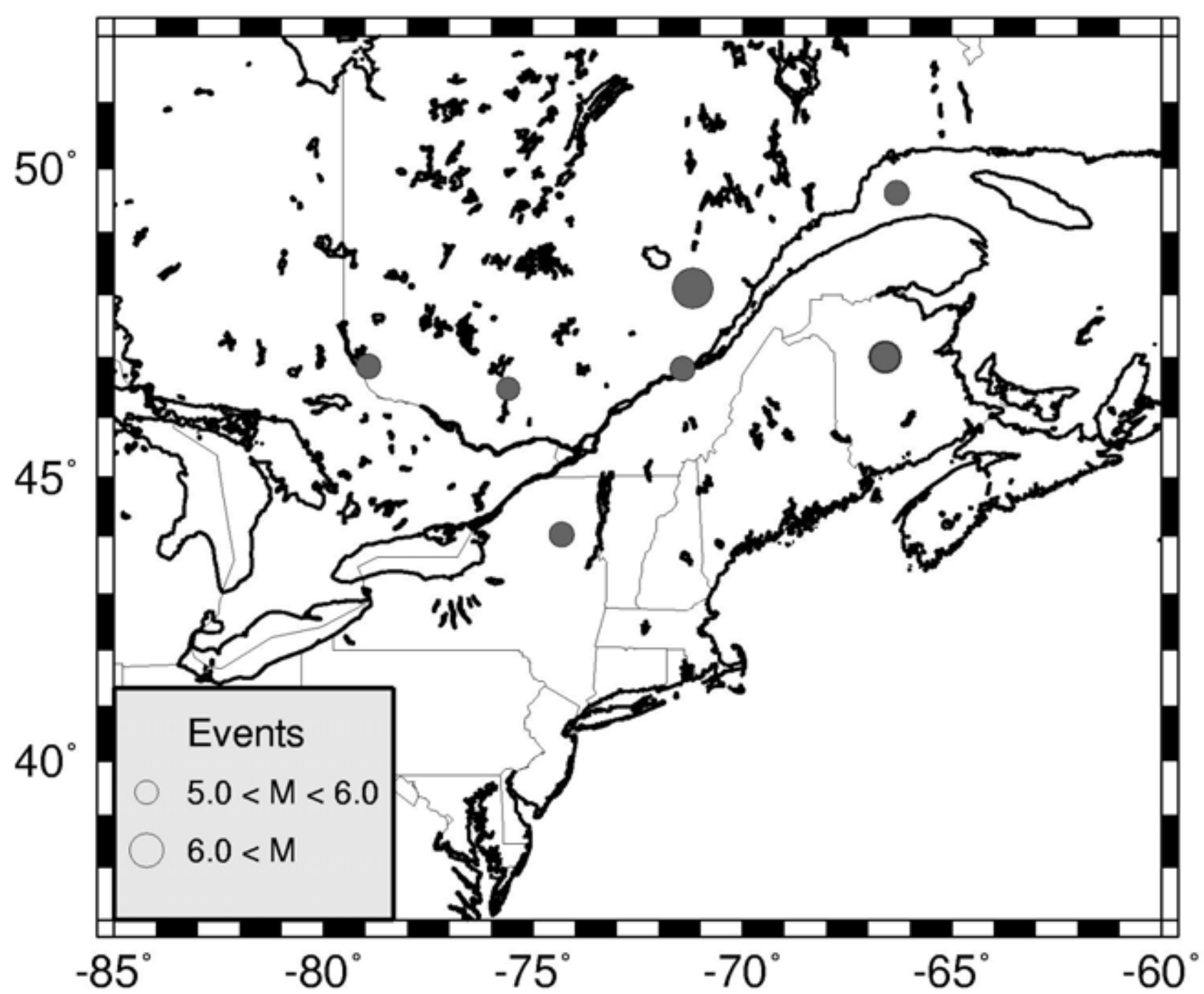

FIG. 2: Large events in eastern Canada, 1900-2001.

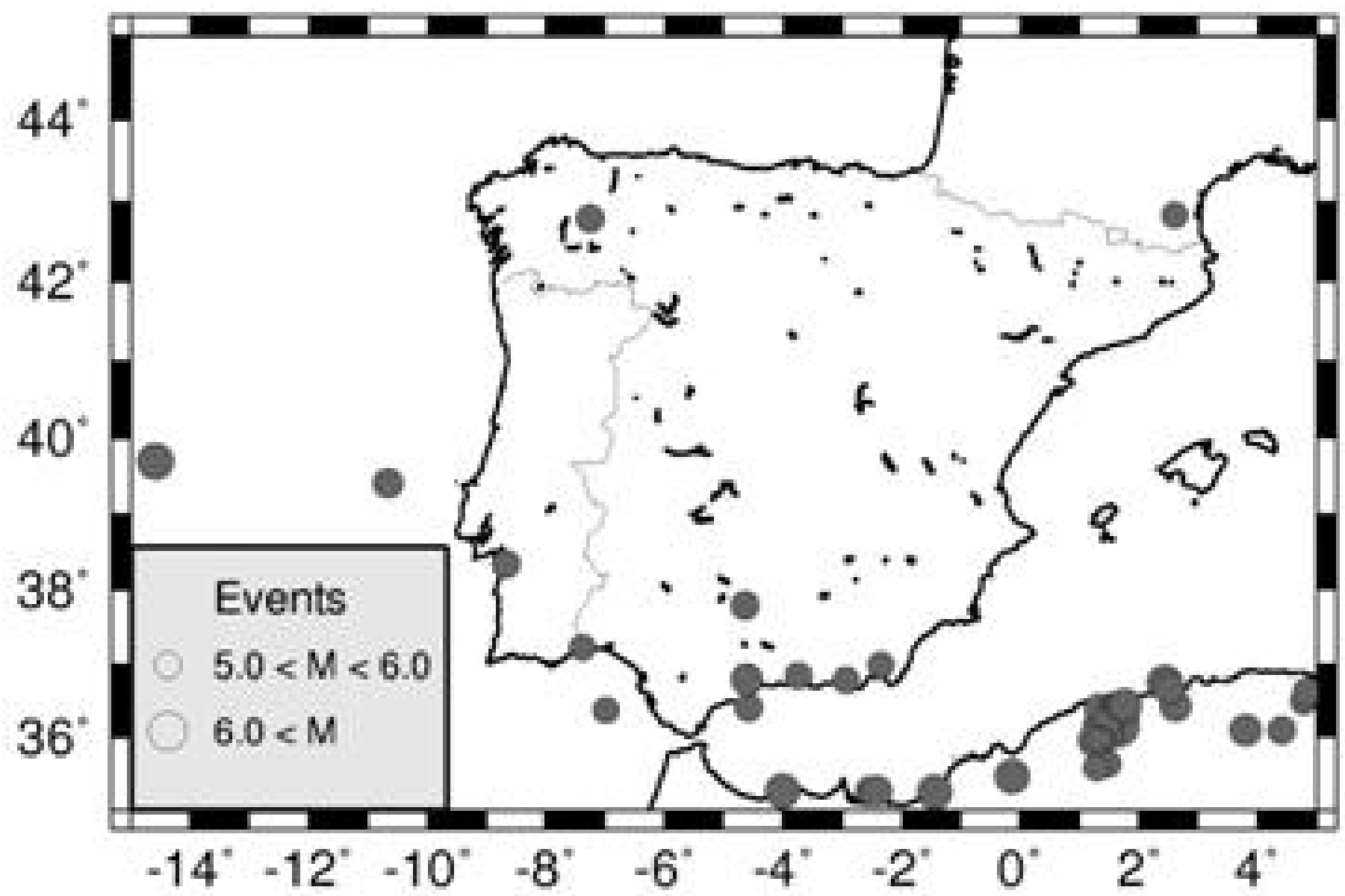

FIG. 3: Large events in Spain, 1970-2001.

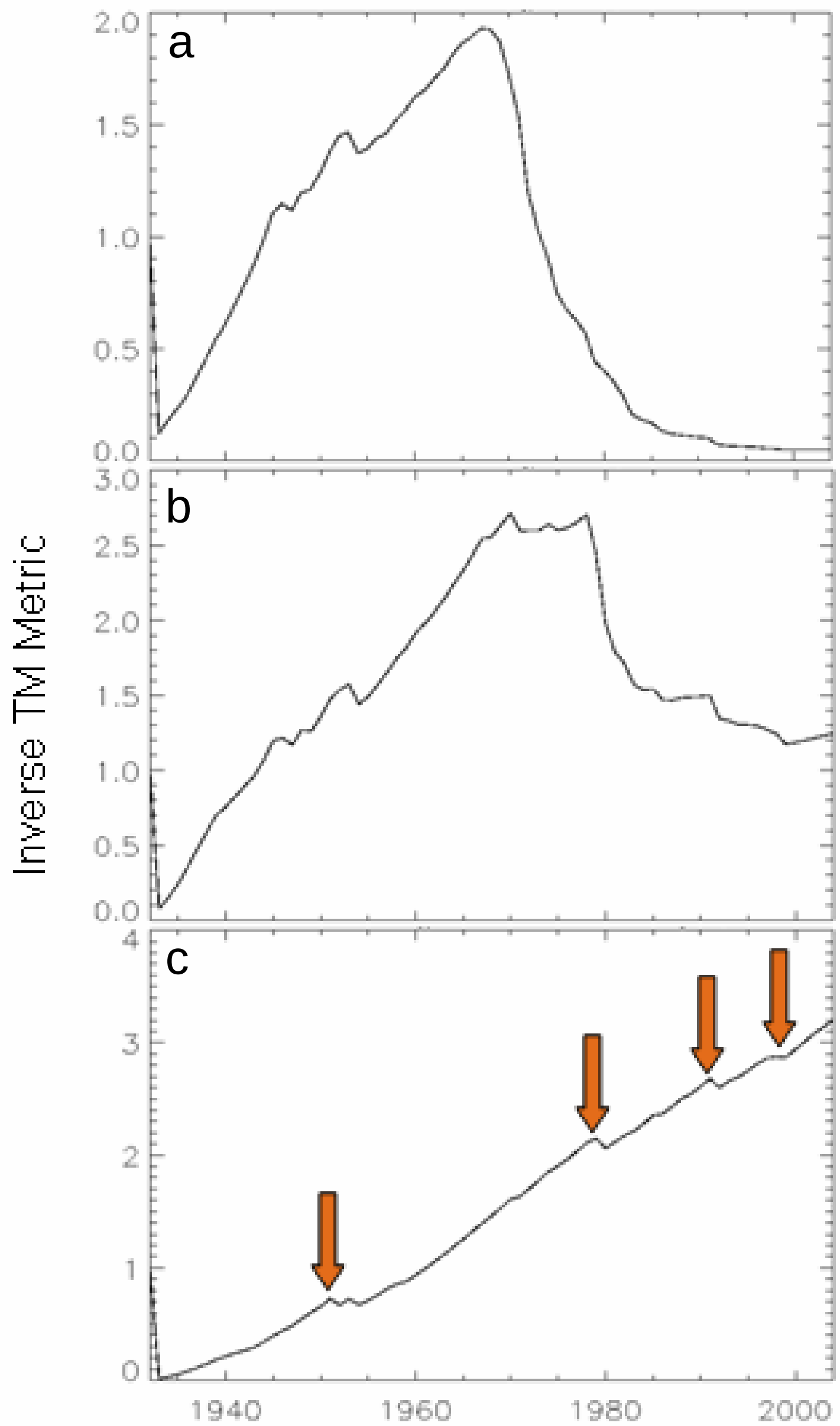

FIG. 4: Plots of the inverse TM metric, normalized to the initial value, for California, for a boxsize of 0.1 degree and a) all M ≥ 2.0, b) all M ≥ 3.0, and c) all M ≥ 4.0.

**Table 1: Major earthquakes in California, 1932-2004**

(compiled from SCEC and USGS data sources)

| Event | Magnitude | Year |
|---|---|---|
| Long Beach | 6.4 | 1933 |
| Kern County | 7.5 | 1952 |
| Parkfield | 6.0 | 1966 |
| Borrego Mountain | 6.5 | 1968 |
| San Fernando | 6.6 | 1971 |
| Imperial Valley | 6.4 | 1979 |
| Coalinga | 6.5 | 1983 |
| N. Palm Springs | 6.0 | 1986 |
| Superstition Hills | 6.6 | 1987 |
| Loma Prieta | 7.0 | 1989 |
| Landers | 7.3 | 1992 |
| Northridge | 6.7 | 1994 |
| Hector Mine | 7.1 | 1999 |
| Parkfield | 6.0 | 2004 |

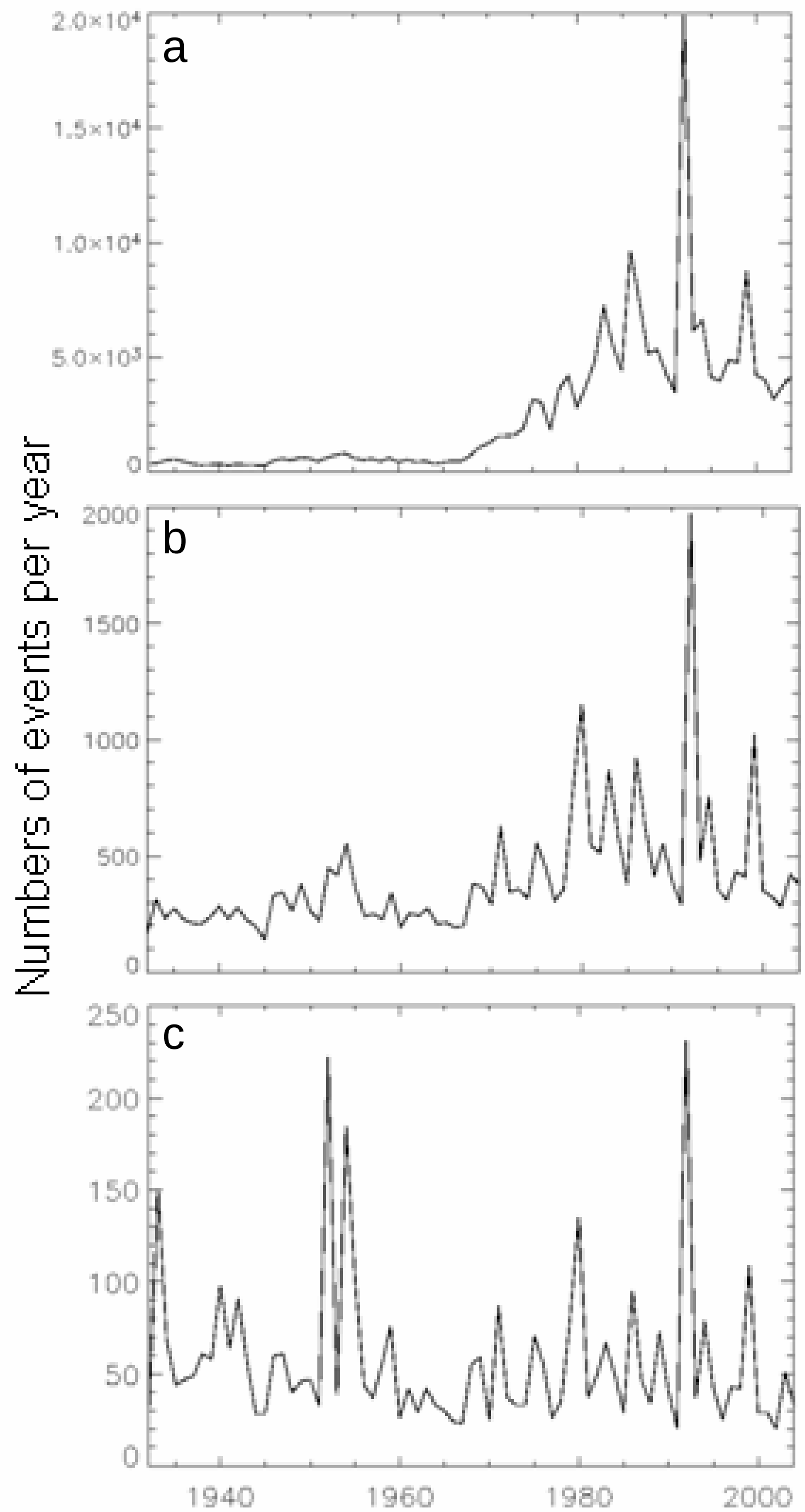

FIG. 5: Count of the number of events per year in California for a) all M ≥ 2.0, b) all M ≥ 3.0, and c) all M ≥ 4.0.

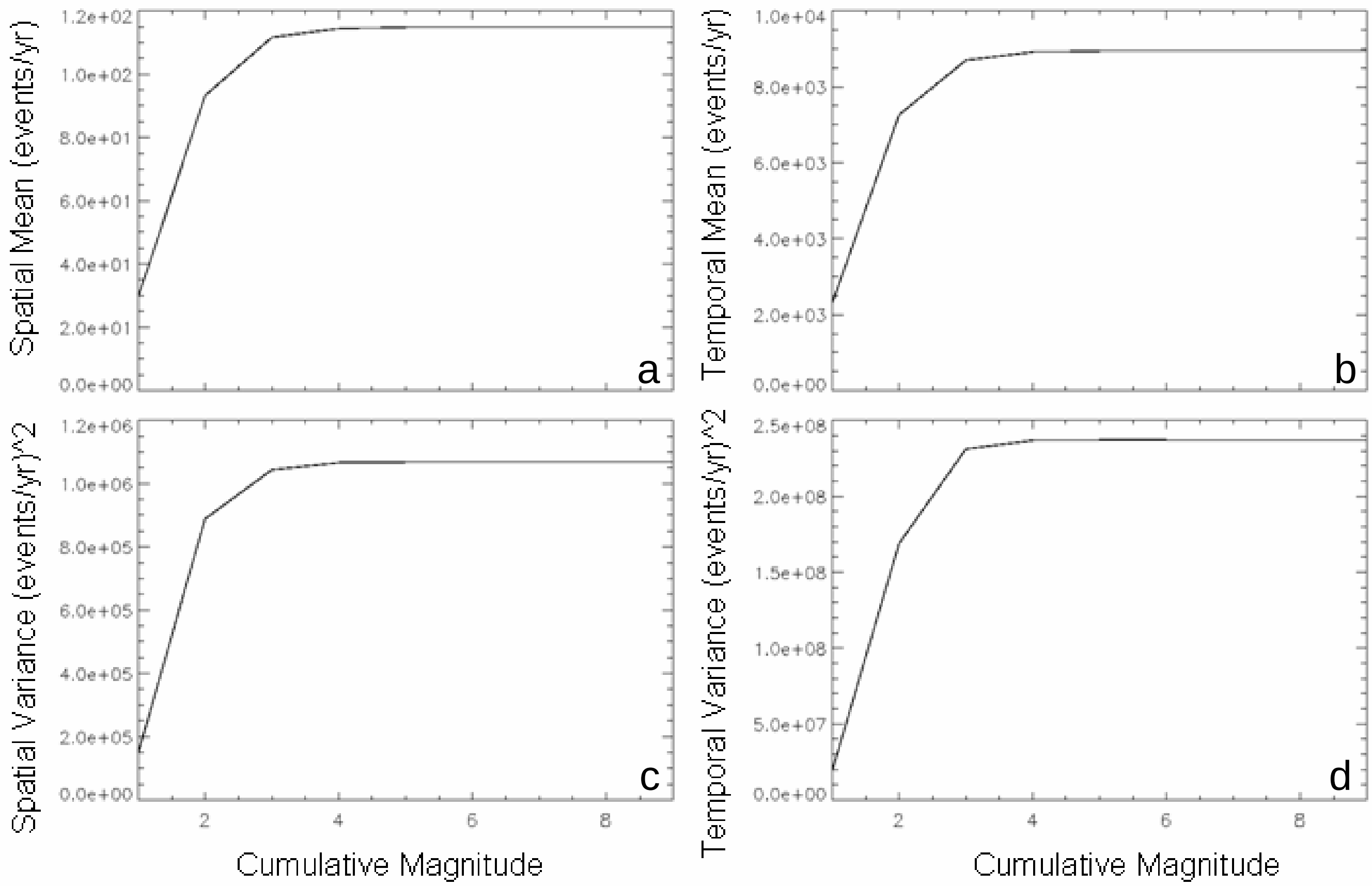

FIG. 6: Statistical quantities for California seismicity for cumulative magnitude (all events less than or equal to that shown). a) Spatial mean, b) temporal mean, c) spatial variance, and d) temporal variance.

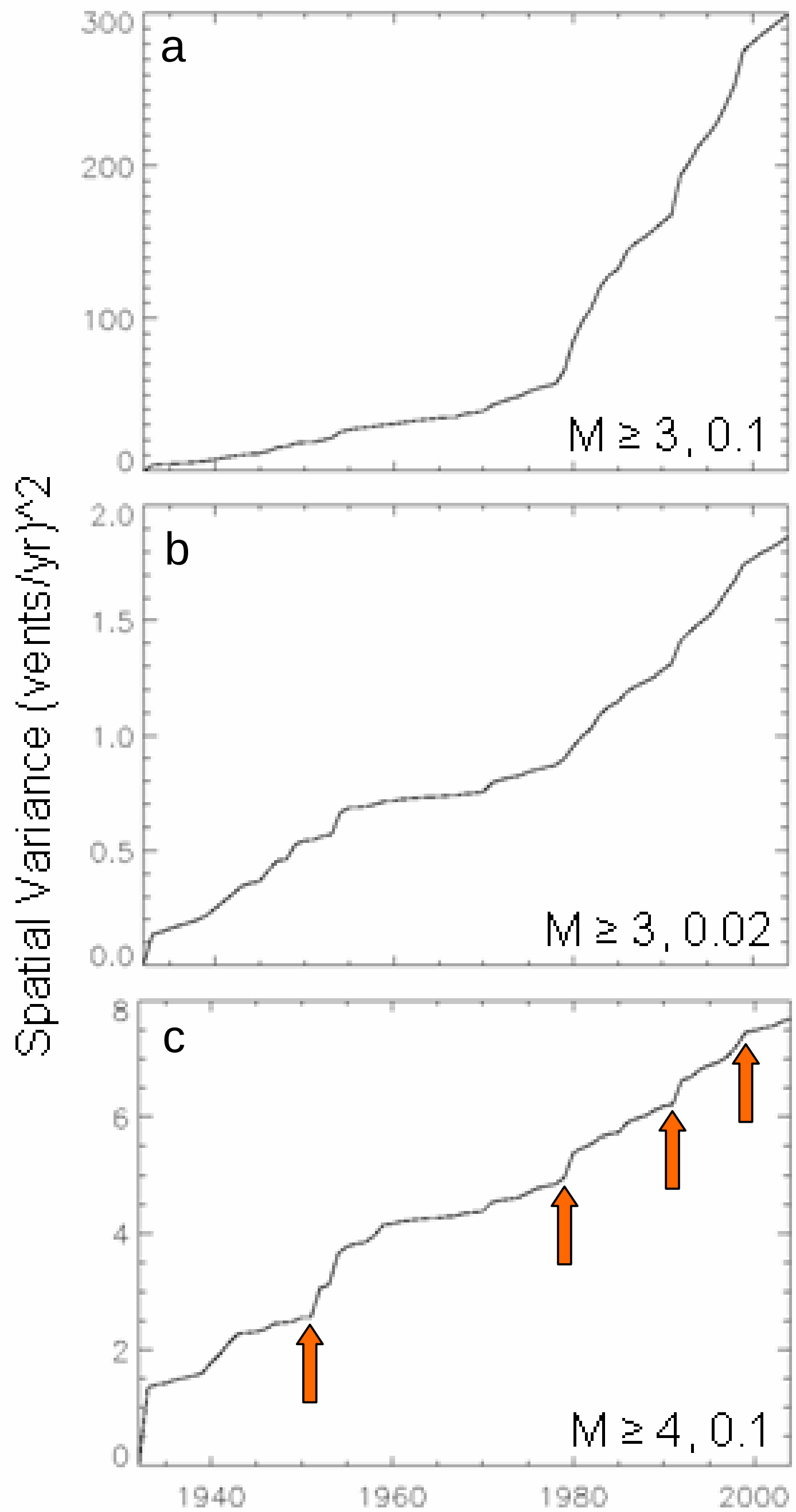

FIG. 7: Spatial variance for California seismicity over time, for events of magnitude greater than or equal to that shown. a) M ≥ 3.0, boxsize = 0.1 degree, b) boxsize = 0.02 degree, M ≥ 3.0, and c) boxsize = 0.1 degree, M ≥ 4.0.

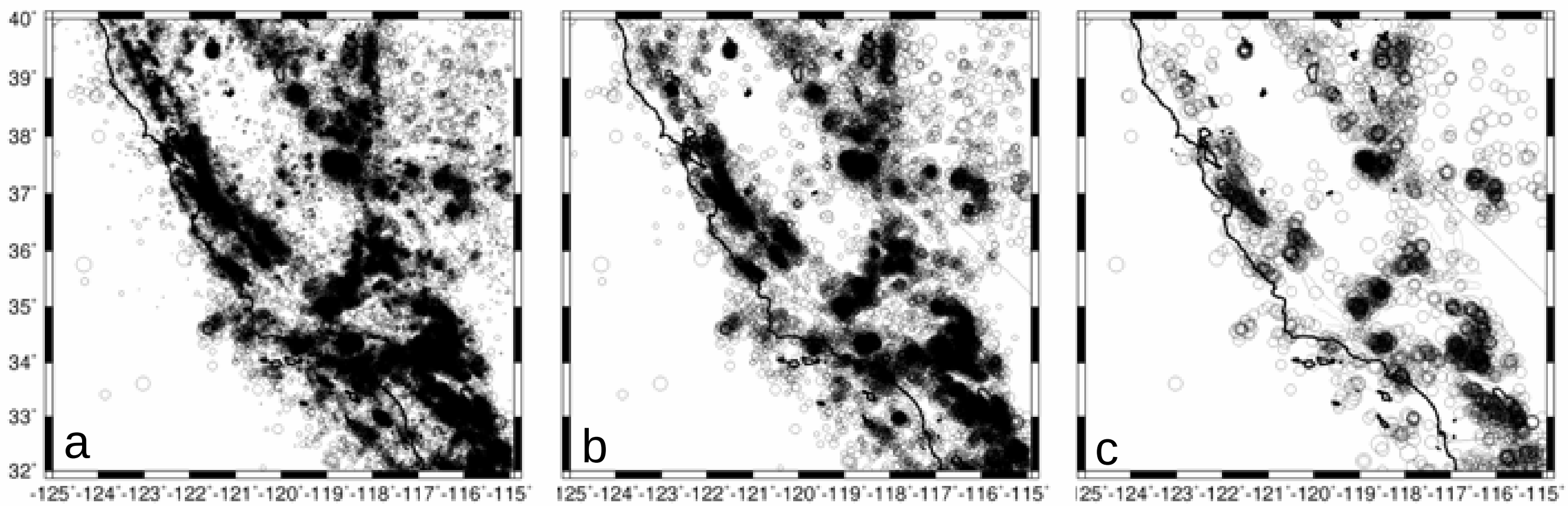

FIG. 8: Plots of California seismicity, 1932-2004, for a) all M ≥ 2.0, b) all M ≥ 3.0, and c) all M ≥ 4.0.

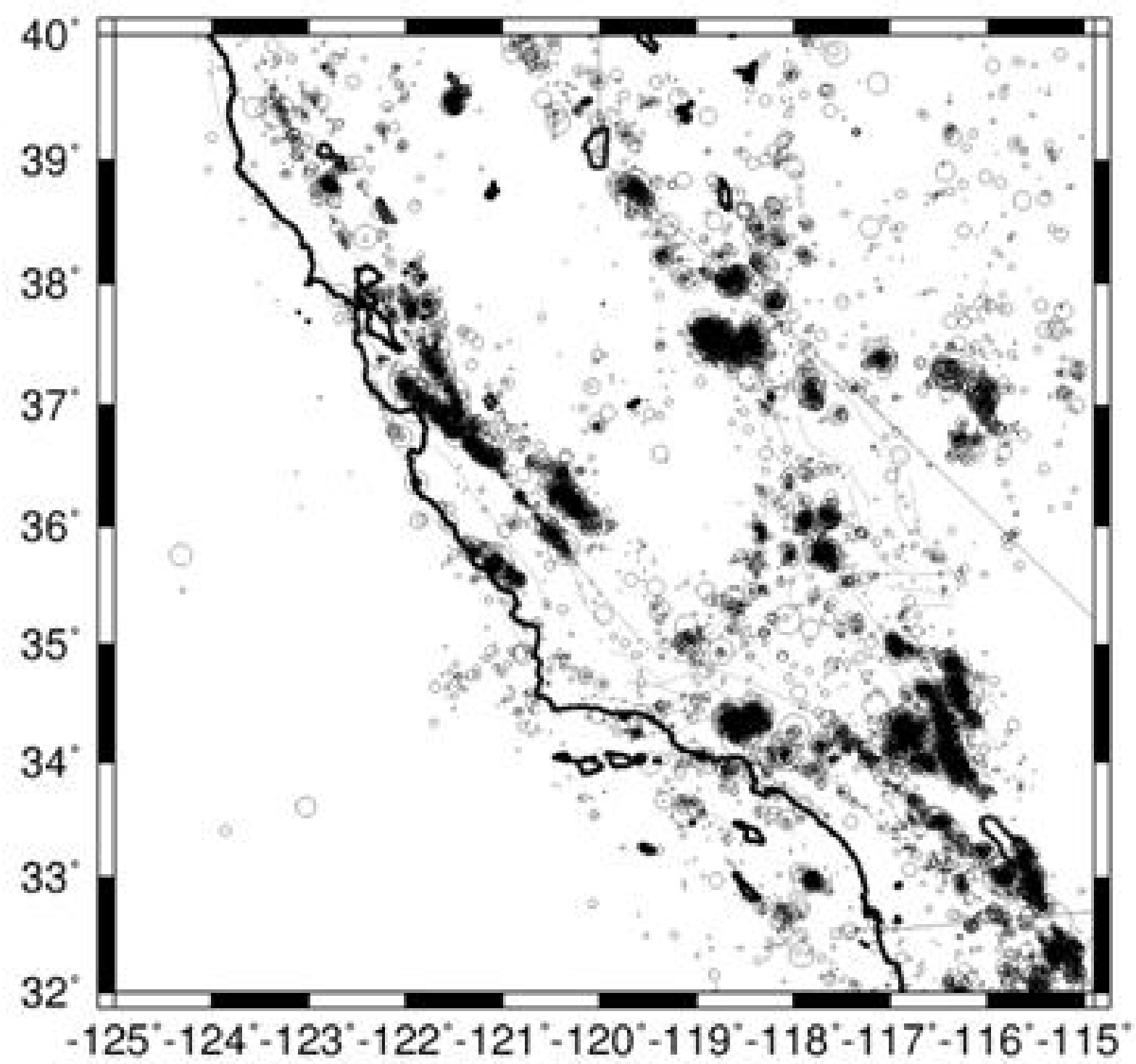

FIG 9: California seismicity, M ≥ 3.0, 1970-2004.

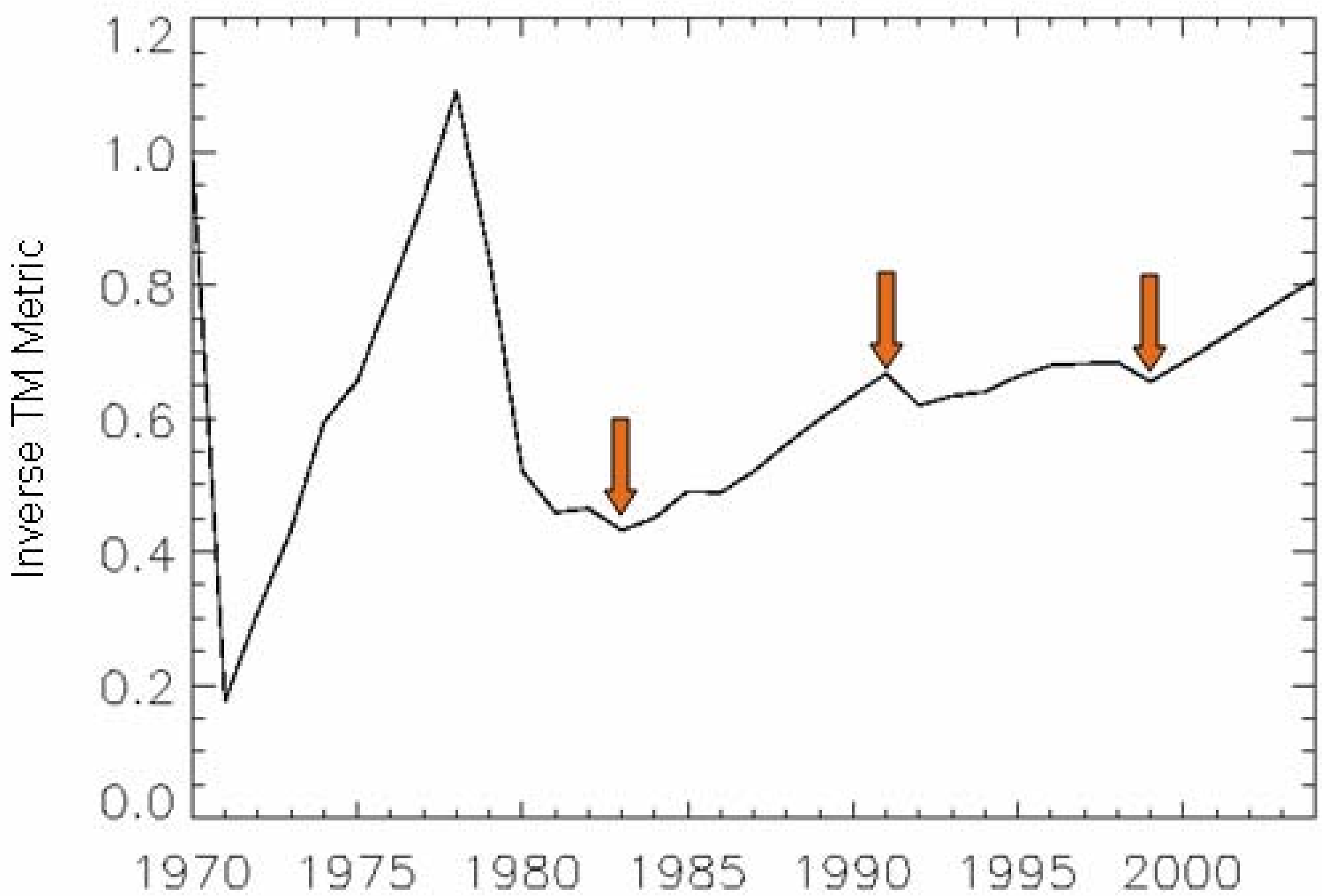

FIG. 10: Inverse TM metric for California seismicity, M ≥ 3.0, for the period 1970-2004, boxsize equal to 0.1 degree.

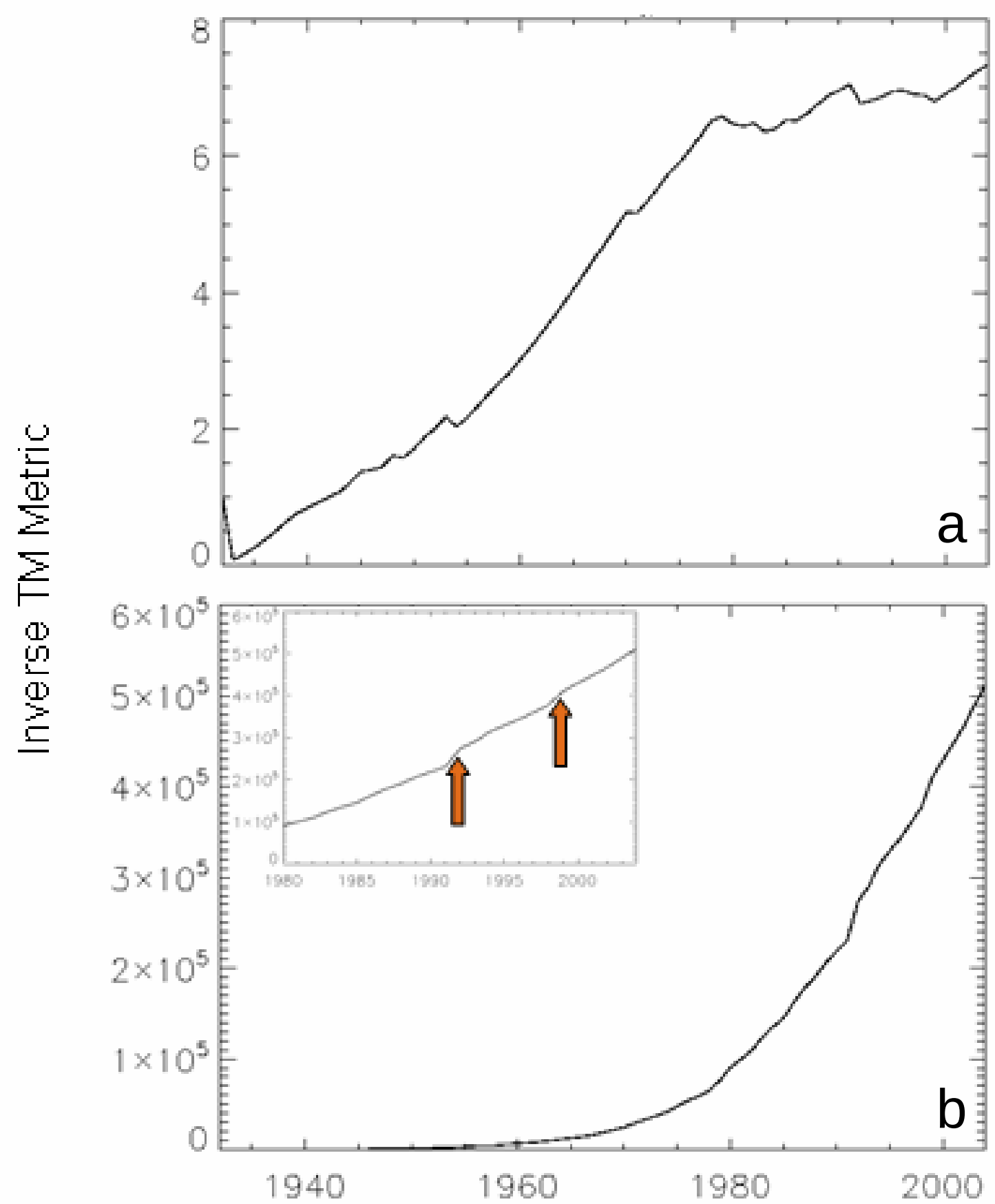

FIG. 11: Inverse TM metric for boxsizes of a) 0.02º (2.2 km), for all boxes and b) 0.0011º (0.12 km), only for those boxes that contain at least one event. Inset shows the inverse metric for 1980 to 2004.

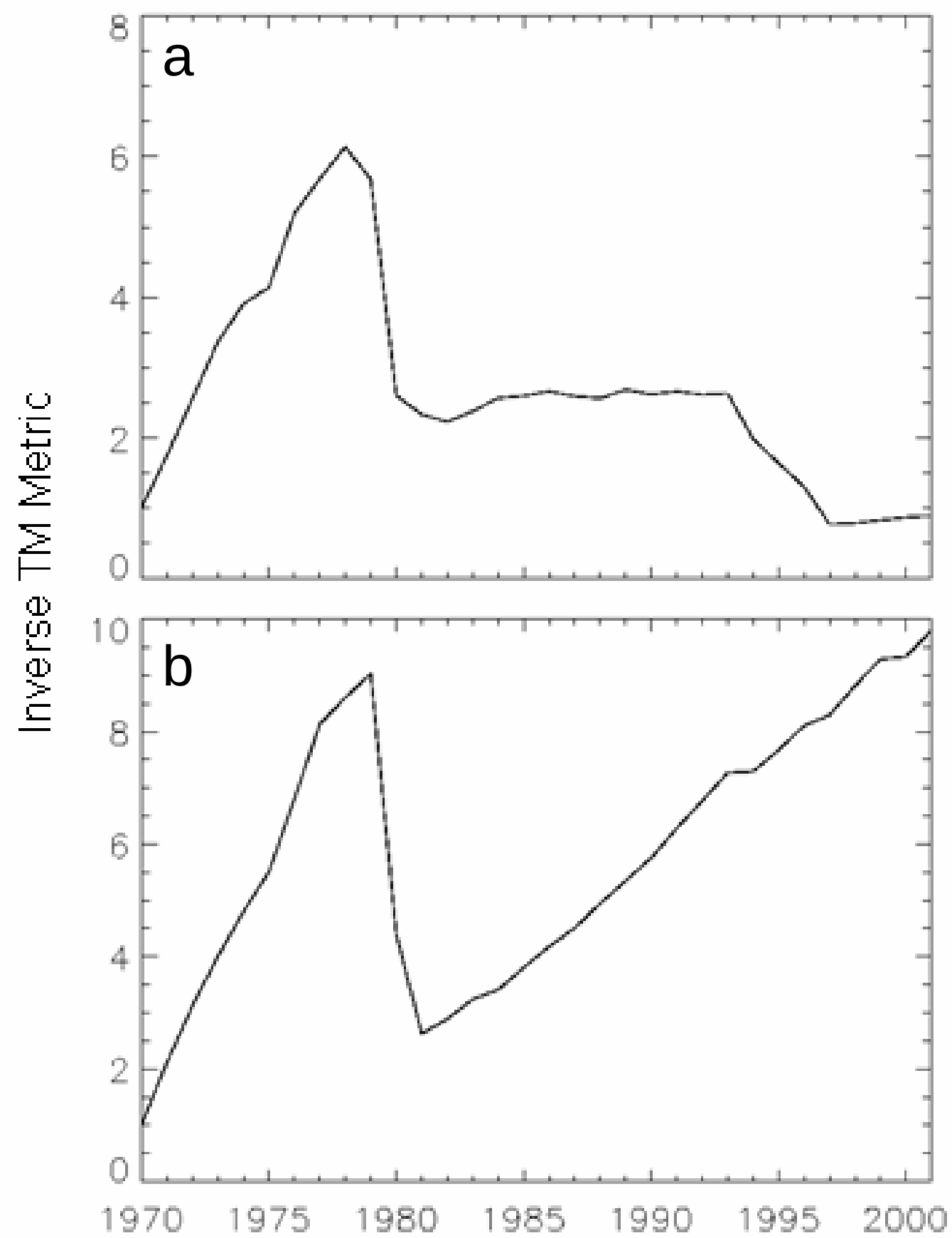

FIG. 12: Inverse TM metric for Iberian peninsula, boxsize = 0.1 degree, a) M ≥ 3.0, and b) M ≥4.0.

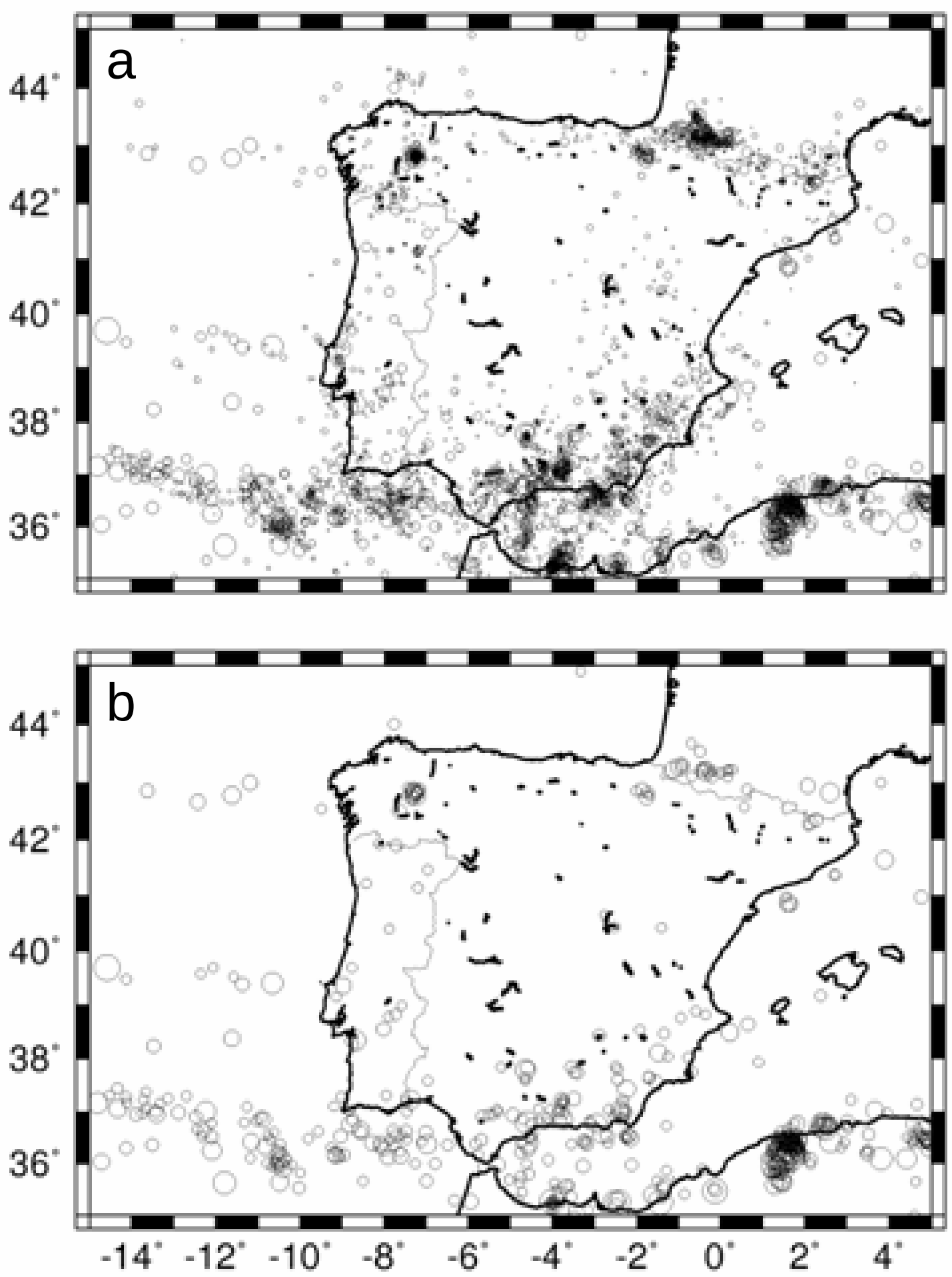

FIG. 13: Seismicity for Iberian peninsula, 1970-2001, a) M ≥ 3.0, and b) M ≥ 4.0.

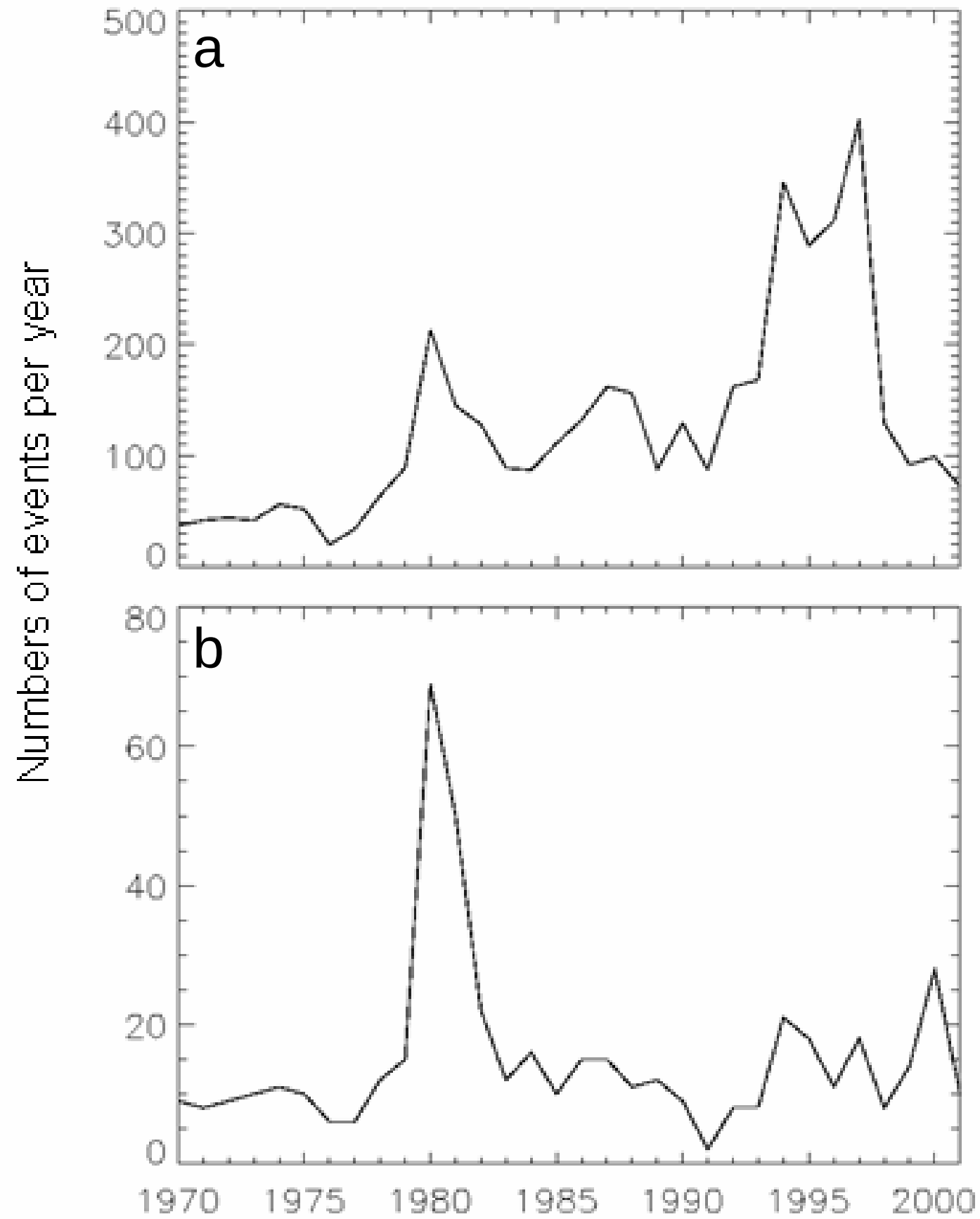

FIG. 14: Numbers of events per year for Iberian Peninsula, a) M ≥ 3.0 and b) M ≥ 4.0.

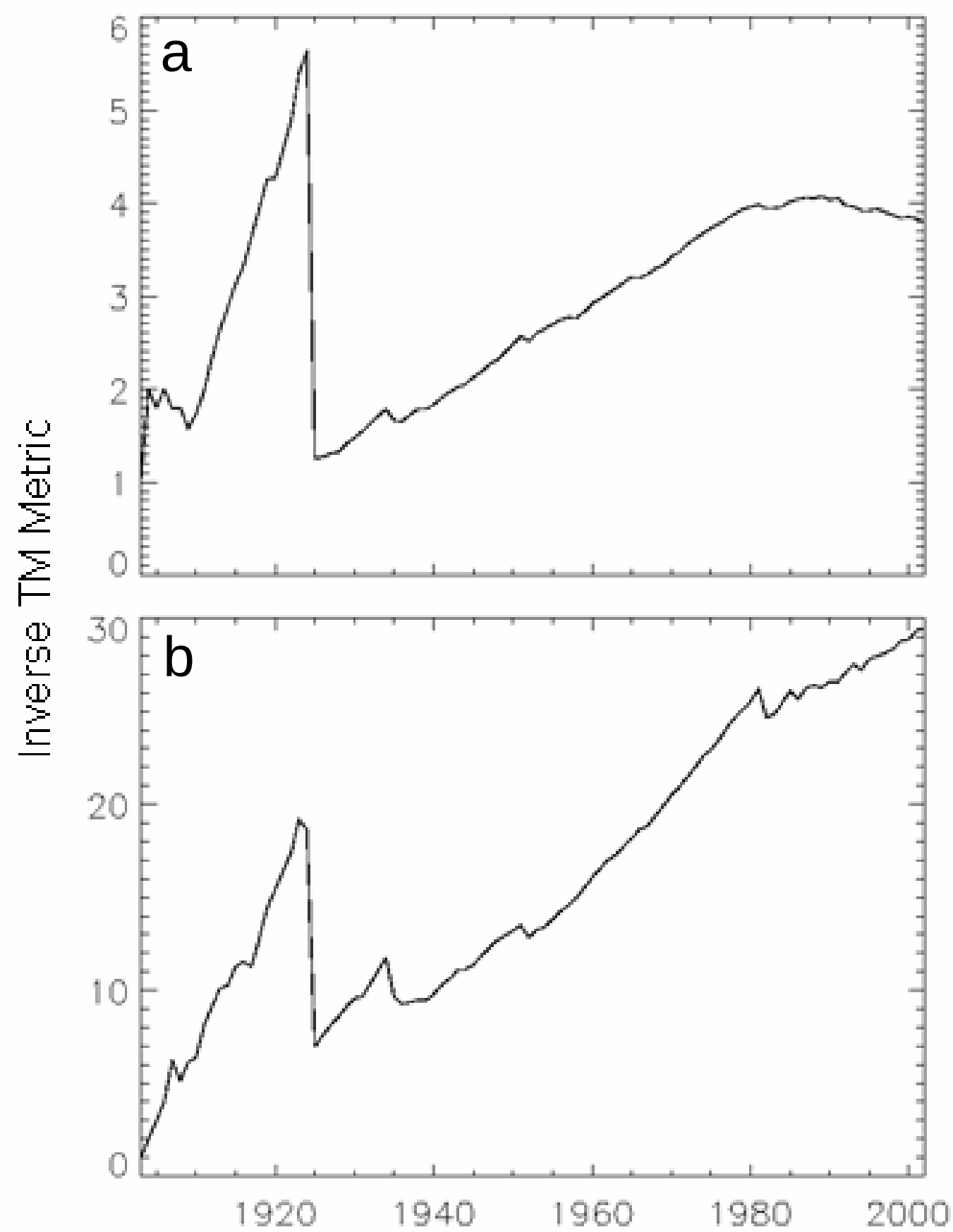

FIG. 15: Inverse TM metric for eastern Canada, 1900-2001 and boxsize = 0.1 degree, a) M ≥ 3.0 and b) M ≥ 4.0.

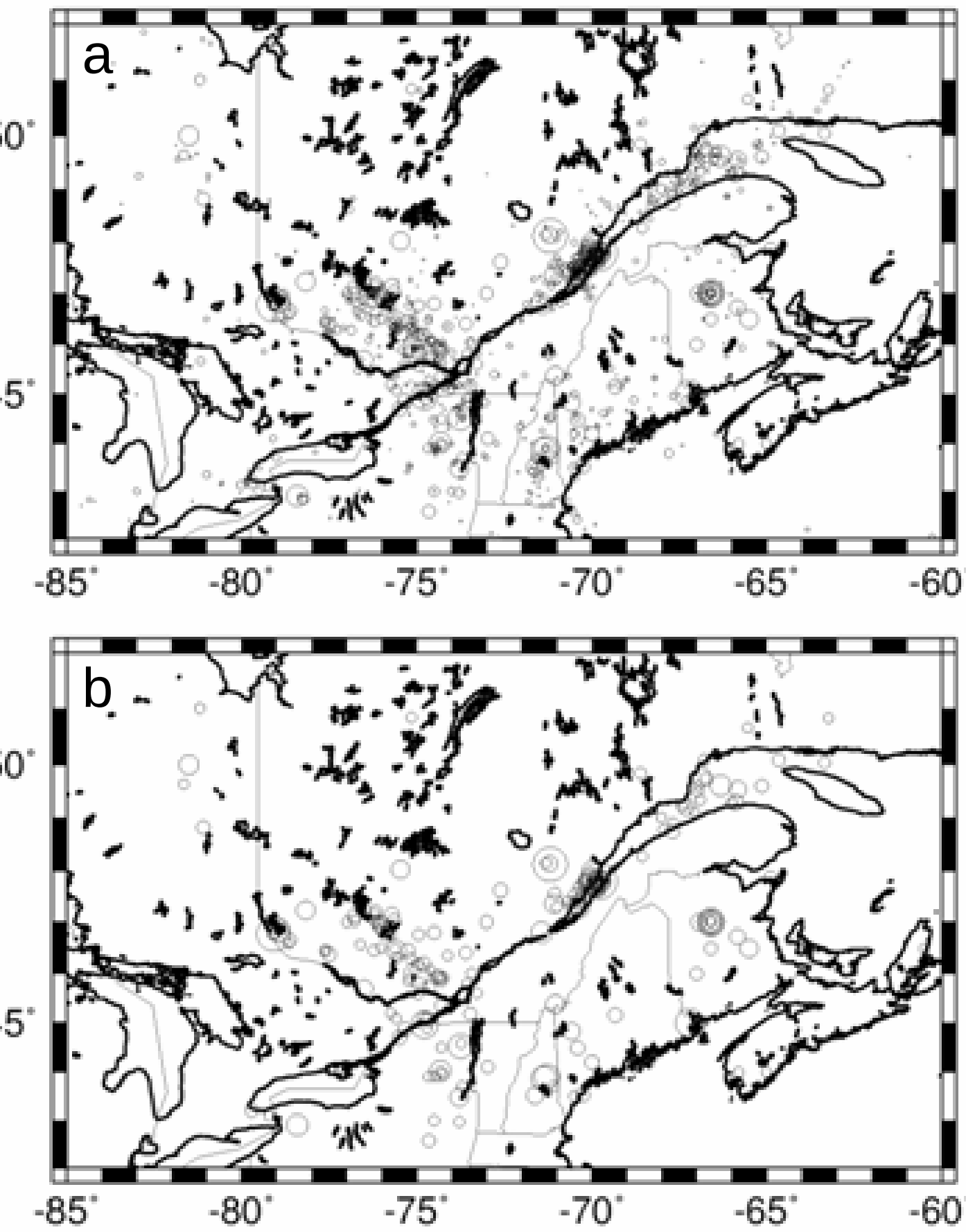

FIG. 16: Seismicity for eastern Canada, 1900-2001, a) M ≥ 3.0 and b) M ≥ 4.0.

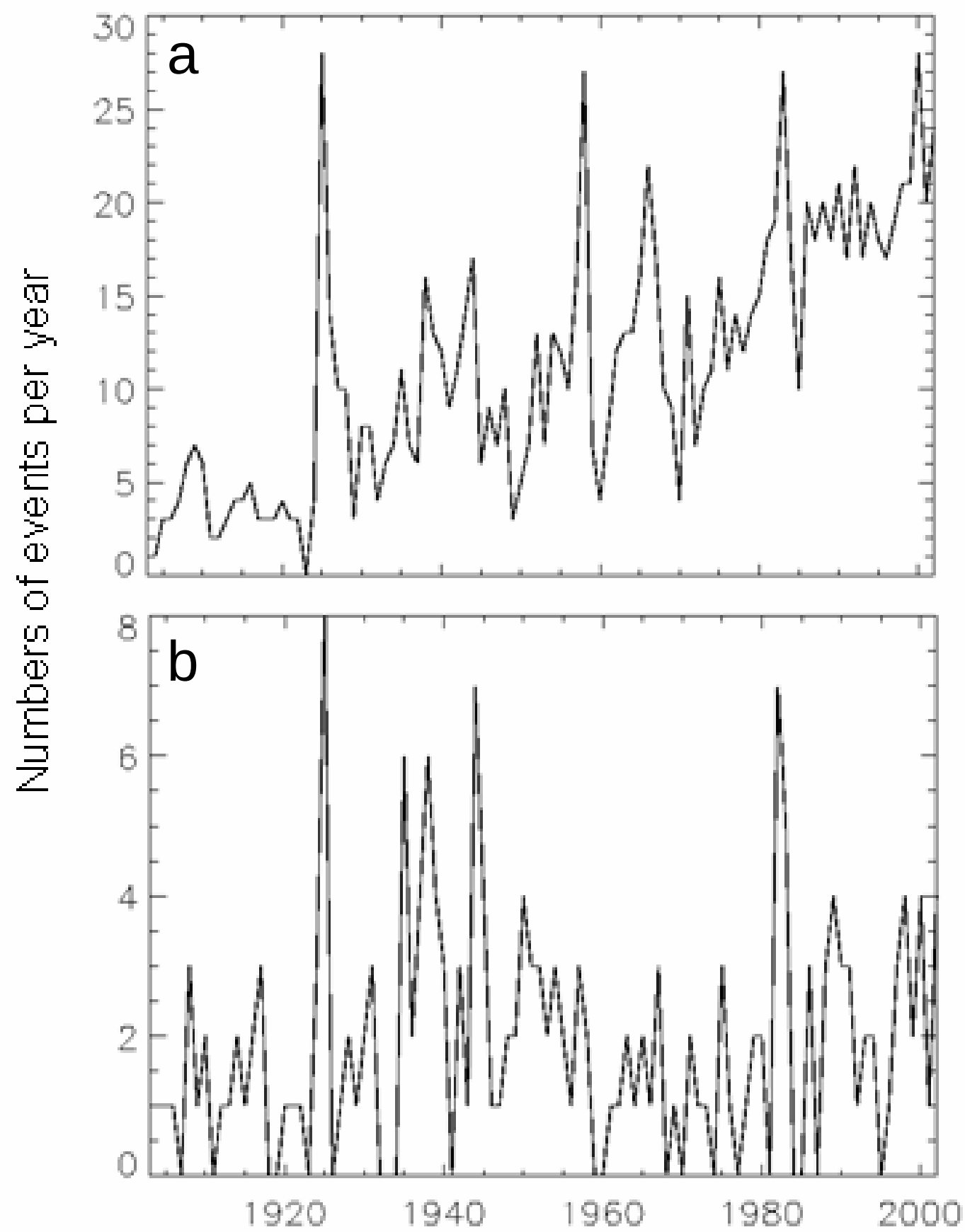

FIG. 17: Number of events per year, 1900-2001 for eastern Canada, a) M ≥ 3.0 and b) M ≥ 4.0.